\documentclass{aa}  

\usepackage{graphicx}
\usepackage{txfonts}
\usepackage[colorlinks=true, linkcolor=blue, filecolor=magenta, citecolor=blue, urlcolor=cyan]{hyperref}
\newcommand{\ff}[1]{\textcolor{blue}{\bf #1}}

\begin{document} 

   \title{Efficient black hole seed formation in low metallicity and dense stellar clusters 
   with implications for JWST sources}


    \author{M.C. Vergara\inst{1}\thanks{E-mail: Marcelo.C.Vergara@uni-heidelberg.de (MV)}
    \and A. Askar \inst{2}\thanks{E-mail: askar@camk.edu.pl (AA)}   
   \and F. Flammini Dotti\inst{3,4,1}
   \and D.R.G. Schleicher\inst{5,6}
   \and A. Escala\inst{7}
   \and R. Spurzem\inst{1,8,9}
   \and \\ M. Giersz \inst{2}
   \and J. Hurley \inst{10,11}
   \and M. Arca Sedda \inst{12,13,14,15}  
   \and N. Neumayer\inst{16}
   }
   
\institute{Astronomisches Rechen-Institut, Zentrum für Astronomie, University of Heidelberg, Mönchhofstrasse 12-14, 69120, Heidelberg, Germany 
\and Nicolaus Copernicus Astronomical Center, Polish Academy of Sciences, Bartycka 18, 00-716 Warsaw, Poland
\and Department of Physics, New York University Abu Dhabi, PO Box 129188 Abu Dhabi, UAE
\and Center for Astrophysics and Space Science (CASS), New York University Abu Dhabi, PO Box 129188, Abu Dhabi, UAE
\and Dipartimento di Fisica, Sapienza Università di Roma, Piazzale Aldo Moro 5, 00185 Rome, Italy
\and Departamento de Astronom\'ia, Facultad Ciencias F\'isicas y Matem\'aticas, Universidad de Concepci\'on, Av. Esteban Iturra s/n Barrio Universitario, Casilla 160-C, Concepci\'on, Chile
\and Departamento de Astronomía, Universidad de Chile, Casilla 36-D, Santiago, Chile
\and National Astronomical Observatories and Key Laboratory of Computational Astrophysics, Chinese Academy of Sciences, 20A Datun Rd., Chaoyang District, Beijing 100012, China
\and Kavli Institute for Astronomy and Astrophysics, Peking University, Yiheyuan Lu 5, Haidian Qu, 100871, Beijing, China
\and OzGrav: The ARC Centre of Excellence for Gravitational Wave Discovery, Hawthorn, VIC 3122, Australia 
\and Centre for Astrophysics and Supercomputing, Department of Physics and Astronomy, John Street, Hawthorn, Victoria, Australia 3122 
\and Gran Sasso Science Institute, Viale F. Crispi 7, I--67100 L'Aquila, Italy 
\and Physics and Astronomy Department Galileo Galilei, University of Padova, Vicolo dell'Osservatorio 3, I--35122, Padova, Italy 
\and INFN – Laboratori Nazionali del Gran Sasso, 67100 L’Aquila, (AQ), Italy 
\and INAF – Osservatorio Astronomico d’Abruzzo, Via M. Maggini snc, 64100 Teramo, Italy
\and Max-Planck-Institut für Astronomie, Königstuhl 17, 69117 Heidelberg, Germany
   }
   
\titlerunning{Black hole seed formation in low metallicity and dense stellar clusters}
\authorrunning{Vergara et al.}

   \date{Received September 15, XXXX; accepted March 16, YYYY}

 
  \abstract
   {Recent observations with the James Webb Space Telescope (JWST) have revealed the presence of young massive clusters (YMCs) as building blocks of the first galaxies during the first billion years of the Universe. They are not only important constituents of the galaxies, but also potential birth places of very massive stars (VMSs) and black hole (BH) seeds. }
   {In this paper, we explore the stellar dynamics in extremely dense clusters with an initial half-mass density of $\rho_{h} \gtrsim 10^8\,\rm M_\odot\,pc^{-3}$ at very low metallicity. These densities are roughly comparable to some of the densest clusters found by JWST. Our detailed $N$-body and Monte Carlo simulations, which include stellar evolution, show that the formation of VMSs through collisions is unavoidable, with the resulting final masses reaching $\sim 5 \times 10^3$ to $4 \times 10^4\,\rm M_\odot$. These simulations  serve to verify the hypothesis that there is a critical mass scale at which collisions in the system become very efficient and thus VMSs and potentially intermediate-mass BHs (IMBHs) can form.}
   {We use \textsc{nbody6++gpu} and  \textsc{MOCCA}, including the latest updates of the stellar evolution routines SSE/BSE, along with specific routines to handle the formation and dynamical evolution of VMSs.}
   {We show that dense star clusters rapidly form VMSs due to constant stellar bombardment. The VMSs eventually collapse and form a BH seed with masses ranging from a few $10^{3}$ to a few $10^{4}\,\rm M_\odot$ in less than $4\,$Myr.}
   {We discover a critical mass-density threshold in star clusters, beyond which the latter experience several runaway collisions, leading to the formation of massive BH seeds. Considering the ratio of stellar mass to critical mass for typical YMCs detected via JWST, we expect efficiencies in the range up to $10\%$ for the so far detected clusters, thus corresponding to expected BH masses up to $10^5\,\rm M_\odot$in case of their formation via collisions, we predict a relation for the BH mass that follows the shape of $\log(M_{BH}~/\rm~M_\odot)=-0.76 + 0.76 \log(M~/\rm~M_\odot)$. As a side product, the frequent formation of VMS may naturally explain the high amount of nitrogen found in galaxies at high redshift.}


   \maketitle
%

\section{Introduction}

Recently, the \textit{James Webb} Space Telescope (\href{https://webb.nasa.gov}{JWST}), has observed enigmatic "little red dots" (LRDs) \citep{Kokorev2024, Akins2025}, which are interpreted as either SMBHs heavily obscured by dust or intensely star-forming dusty galaxies \citep{Matthee2024, Napolitano2024}. If these objects host active galactic nuclei (AGNs), it would indicate that most high-redshift galaxies indeed harbour central SMBHs \citep{Greene2024}. On the other hand, if dominated by star formation, LRDs could represent the most compact stellar systems ever observed \citep{Guia2024}.


As mentioned above, the currently enigmatic LRDs could well represent the most high redshift galaxies harboring a central SMBH \citep{Greene2024}, and/or the most compact stellar systems ever observed with central densities around $10^8~\rm M_\odot/pc^3$ \citep{Guia2024}. Some models also propose tidal disruption events (TDEs) of surrounding stars falling into an SMBH as potential explanations of their emission \citep{Bellovary2025}. The estimated stellar to BH mass ratios of these objects are approximately two to three orders of magnitude higher than those observed in the local Universe, which has led to their classification as "overmassive" SMBHs \citep{Goulding2023, Scoggins2023, Ubler2024, Furtak2024, Juodzbalis2024}. These overmassive SMBHs challenge existing models of BH growth, as they imply that BH grew extremely fast in the early Universe.

LRDs are not the only mysterious objects observed by JWST; the presence of young massive clusters (YMCs) in the early Universe also challenges current models of cosmic evolution. These stellar systems are extremely massive and dense, and their presence at high redshift implies that they must have formed within the first few hundred million years after the Big Bang \citep{Vanzella2022b, Vanzella2022,Vanzella2022c, Vanzella2023, Adamo2024a, Mowla2024}. Gas-rich dwarf galaxy mergers at high redshift have been suggested as a possible channel to form these massive and compact stellar systems \citep{renaud2015, lahen2020a, lahen2020b, Lahen2025}. These highly dense stellar environments are ideal places for runaways collision to occur and form a VMS \citep{Spitzer1966, Spitzer1967, Sanders1970, Lee1987, Quinlan1990, Gurkan2004, Freitag2006b, Freitag2006a, Gierszetal2015, Vergara2023, Vergara2024, Vergara2025, Rantala2025, Rantala2025b} and also are suitable places for chemical self-enrichment \citep{Vink2018a}. These primordial dense stellar systems and their capability to form VMSs can be a possible explanation for other mysteries observations by the JWST, such as the high nitrogen-to-oxygen ratio in galaxies at high redshift \citep{Charbonnel2023a}, which is around four times larger than in the solar vicinity \citep{Cameron2023b}. Some examples of these galaxies with high nitrogen abundances are GN-z11 at redshift $10.6$ \citep{Bouwens2010a, Tacchella2023a, NageleUmeda2023, Maiolino2024}, CEERS-1019 at $z=8.679$ \citep{Larson2023, Marques-Chaves2024} and the most distant one to the date MOM-z14 at $z=14.44$ \citep{Naidu2025}. 


To explain the formation of VMSs and SMBHs, several scenarios had been proposed in the literature \citep{Rees1978, Rees1984}. One prominent scenario involves the remnants of Population III (Pop III) stars, which form in metal-free clouds and accrete mass onto their cores \citep{Bond1984, Omukai1998, Madau2001, Volonteri2003, Tan2004, Ricarte2018, Mestichelli2024, Reinoso2025, Solar2025}. Pop III stars under rapid rotation are potential source of nitrogen enrichment \citep{Tsiatsiou2024, Nandal2024}. Alternatively, the direct collapse of gas clouds \citep{Loeb1994, Begelmanetal2006, Lodato2006, Chon2024}, possibly through a quasi-star phase \citep{Begelman2010}, has been suggested as another viable pathway. Stellar dynamics have been proposed as a possible pathway \citep{Portegies1999, Portegies2002, Reinoso2018, Reinoso2020, Alister-Seguel2020, Vergaraetal2021, Vergara2023, Vergara2024, Vergara2025}. 

The early attempts to explain the high luminosities observed in galactic centers relied on analytical models.  \citet{Spitzer1966, Spitzer1967} demonstrated that luminosities of $\sim 10^{43} ~ \mathrm{erg\,s^{-1}}$ can arise in dense star clusters of $10^8~ \mathrm{M_\odot}$ within a radius of $\sim 0.1~ \mathrm{pc}$. Similarly, \citet{Sanders1970} showed that VMS could form in stellar systems with short relaxation times. VMS formation via collisions was first predicted using Fokker-Planck models, which indicated that a massive star naturally forms at the cluster core due to the deep gravitational potential \citep{Lee1987}. Later, \citet{Quinlan1990} proposed that dense galactic nuclei could host stars with masses of thousands of solar masses, potentially seeding SMBHs. Monte Carlo simulations by \citet{Gurkan2004} revealed that dense clusters undergo core collapse faster than the lifetime of a massive star, enabling sustained mass growth before collapse. In particular, clusters with over a million stars exhibit high collision rates, facilitating the formation of $\sim 10^3 ~ \mathrm{M_\odot}$ stars that may collapse into IMBHs \citep{Freitag2006a, Freitag2006b,Gierszetal2015}.  Direct \textit{N}-body simulations have also produced massive stars of hundreds of solar masses \citep{Portegies1999,portegies-zwart2004a}. The first million-particle $N$-body simulation, {\sc dragon} \citep{Wangetal2016}, marked a significant milestone. Subsequent studies, including the {\sc dragon-II} simulations by \citet{ArcaSeddaetal2023a, ArcaSeddaetal2024a, ArcaSeddaetal2024b}, have investigated the formation of IMBHs in star clusters with central densities of approximately $10^5~\mathrm{M_\odot\, pc^{-3}}$. \citet{Escala2021} further argued that stellar systems become prone to global instability, thus conducive to massive object formation, if their average collision timescale is comparable to or shorter than their age, and discussed that primordial stellar clusters are ideal candidates to suffer such physical process. \citet{Vergara2023} has demonstrated with equal-mass star models that above a certain critical mass, defined when the average collision timescale is equal to their age, collapse becomes inevitable and massive objects can form in clusters with central densities up to $10^{10} ~ \mathrm{M_\odot\,pc^{-3}}$. This finding was expanded to different stellar systems, including not only NSCs, but also Globular Clusters (GCs) and Ultra-Compact Dwarf Galaxies (UCDs), covering different initial conditions, stellar initial mass functions, and evolutionary paths \citep{Vergara2024}. Using the critical mass framework, \citet{Liempi2025} reproduced the observed mass distribution of NSCs and SMBHs with semi-analytical models. Recently, \citet{Vergara2025} conducted one of the most computationally intensive million-particle simulations to date, with a central density of $> 10^7 ~ \mathrm{M_\odot\,pc^{-3}}$, forming an IMBH of $> 5 \times 10^4 ~ \mathrm{M_\odot}$ within $5\,$Myr. Another scenario involves relativistic clusters, where stellar BH dynamics may facilitate SMBH formation \citep{Shapiro1985, Lupi2014, Kroupaetal2020, Gaete2024, Bamber2025}.

This paper is structured as follows: Section 2 presents the methodology and introduces the initial conditions. Section 3 presents the results. Finally, section 4 provides a summary of the conclusions along with a discussion of theoretical aspects.

\section{Methodology} \label{section_meth}

For this study, we performed simulations with the direct \textit{N}-body code \href{https://github.com/nbodyx/Nbody6ppGPU}{\textsc{Nbody6++GPU}} and \textsc{MOCCA} \footnote{Monte Carlo Cluster simulAtor}, both codes have been widely compared \citep{Gierszetal2008, Gierszetal2015, Heggie2014, Wangetal2016,Madridetal2017,Kamlahetal2022a, Vergara2025}, showing excellent agreement in stellar dynamics and individual stellar properties. Both share  stellar and binary evolution recipes based on the SSE/BSE population synthesis codes \citet{Hurleyetal2000, Hurleyetal2002a, Hurleyetal2005} and updates to them by \citet{Banerjeeetal2020,Kamlahetal2022a, SpurzemKamlah2023} and \citet{Vergara2025}.

\subsection{\textsc{Nbody6++GPU}}

\href{https://github.com/nbodyx/Nbody6ppGPU}{\textsc{Nbody6++GPU}}  is a high-precision direct \textit{N}-body code \citep{Spurzem1999, Wangetal2015, SpurzemKamlah2023}. It includes several algorithms to solve the stellar dynamics such as the Kustaanheimo-Stiefel regularization, an algorithm to solve close encounters and to form binaries \citep{Stiefel1965}, the chain regularization \citep{MikkolaAarseth1990, MikkolaAarseth1993}, the $4^{th}$ order Hermite integrator scheme with hierarchical block time-steps \citep{McMillan1986, Hutetal1995, MakinoAarseth1992, Makino1999}, the Ahmad-Cohen neighbour scheme which spatially splits the stellar hierarchy to speed up computational calculations \citep{AhmadCohen1973}, and parallelization and acceleration that allow large number of particles, using GPU \citep{Wangetal2015}, e.g. \citet{Wangetal2016, ArcaSeddaetal2023a, ArcaSeddaetal2024a, ArcaSeddaetal2024b, Vergara2025}. It enables realistic simulations of star clusters. The algorithms behind the treatment of the stellar dynamics are explained in more detail in the review article by \citet{SpurzemKamlah2023} on collisional stellar systems.

\subsection{\textsc{MOCCA}}

\textsc{MOCCA} is a code for simulating the evolution of realistic star clusters \citep{Gierszetal2013,HypkiGiersz2013}. It is based on the Monte Carlo method developed by \citet{Henon1971}, which was subsequently improved by \citet{Stodolkiewicz1982,Stodolkiewicz1986} and \citet{Giersz2001}. This method combines the particle-based approach of direct $N$-body simulations with a statistical treatment of two-body relaxation, allowing for efficient computation of the long-term dynamical evolution of spherically symmetric star clusters. \textsc{MOCCA} includes treatments for important physical processes that drive cluster evolution, including stellar and binary evolution(SSE/BSE as in \textit{N}-body), the effects of a galactic tidal field, and the direct integration of strong dynamical encounters using the FEWBODY code \citep{Fregeauetal2004} for scattering experiments.

\subsection{Initial conditions}

\begin{figure*}[!h]
    \centering    
    \includegraphics[]{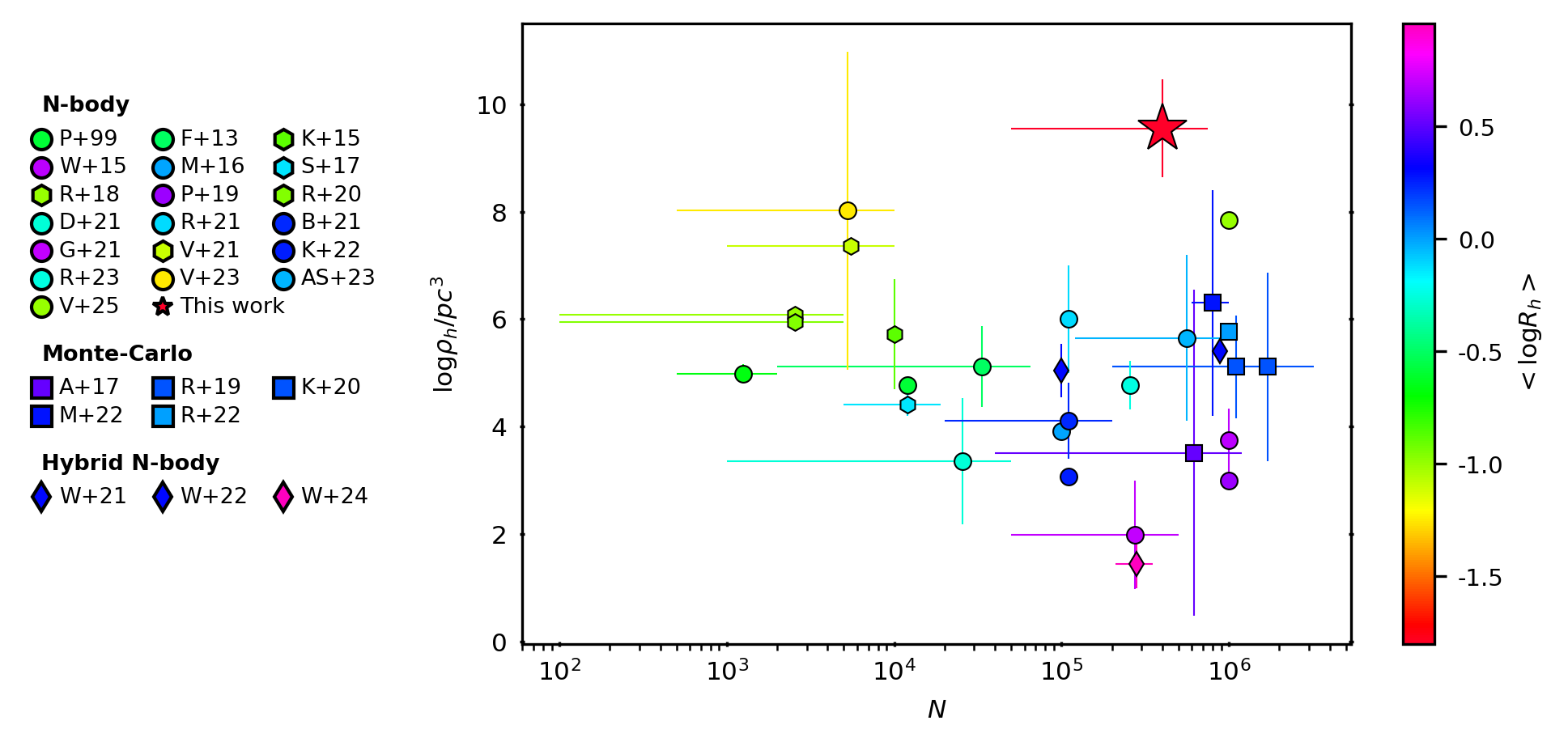}
    \caption{Initial half-mass density $\rho_{h}$, computed at the initial half-mass radius $r_{h}$, is expressed as a function of the initial number of stars $N$. Figure has been reproduced from \citet{ArcaSeddaetal2023a, ArcaSeddaetal2024a, ArcaSeddaetal2024b} and has been adapted to include a color bar with the logarithm of the average half-mass radius. Besides we include others simulations and the new ones presented in this paper.} \label{Rho_vs_N}
\end{figure*}

We modeled 5 isolated clusters using a King density profile \citep{Kingetal1968}, with $W_0=6$, including a Kroupa initial mass function (IMF) \citep{Kroupa2001} with range $M_*=0.08-150~\rm M_\odot$. We varied the number of stars as $N = 5\times10^4, 10^5, 2\times10^5, 5\times10^5, 7.5\times10^5$, we do not include primordial binaries, the cluster half-mass radius varied as $R_h=0.005, 0.01, 0.05\, \rm pc$ and the absolute metallicity is $10^{-4}$ (see Table~\ref{table1-ic}). We use \href{https://github.com/agostinolev/mcluster.git}{\textsc{McLuster}} to generate the initial conditions for the \textit{N}-body and \textsc{MOCCA} codes. Models \texttt{R005N750k} and \texttt{R005N500k} have been motivated by some of the most recent observations with JWST at high redshift, that revealed dense stellar systems with effective radii of $\lesssim 1\,\rm pc$ and masses of the order  $10^6\,\rm M_\odot$ \citep{Vanzella2022b, Adamo2024a}. Models \texttt{R001N250k}, \texttt{R001N100k}, and \texttt{R0005N50k} correspond to a proof of concept to show the expected evolution in a higher density regime.
This work investigates compact, low-particle clusters to study massive object formation in dense stellar systems. Computational constraints limit us to smaller numbers of stars, but still allow us to explore the effect of shorter dynamical timescales.
    

\begin{table}
    \caption{The first column lists the model names, followed by the half-mass radius in the second column, then in the third column is the initial number of stars, the fourth and fifth columns are the mass and central half density of the cluster, respectively and the last column is the absolute metallicity $Z$}
    \begin{tabular}[{0.5\textwidth}]{llcccc}\hline\hline
       Models & $R_{h}$ & $N$ & $M$ & $\rho_{h}$ &$Z$\\
       & $[\mathrm{pc}]$ & $10^5$ & $[10^5\mathrm{M_\odot}]$ &$[\mathrm{M_\odot\, pc^{-3}}]$&\\\hline
       \texttt{R005N750k} & $0.05$ & $7.5$ &$4.38$&$4.18\times 10^8$&$10^{-4}$\\
       \texttt{R005N500k} & $0.05$ & $5$ &$2.93$&$2.79\times 10^8$&$10^{-4}$\\
       \texttt{R001N250k} & $0.01$ & $2.5$ &$1.46$&$1.74\times 10^{10}$&$10^{-4}$\\
       \texttt{R001N100k} & $0.01$ & $1$ &$0.58$&$6.92\times 10^9$&$10^{-4}$\\
       \texttt{R0005N50k} & $0.005$ & $0.5$ &$0.28$&$2.67\times 10^{10}$&$10^{-4}$\\
    \hline\hline    
    \end{tabular}
    \label{table1-ic}
\end{table}
    
In Fig.~\ref{Rho_vs_N}, we display the initial half-mass density $\rho_{\mathrm{h}}$ of the star cluster model, evaluated at the initial half-mass radius $r_{\mathrm{h}}$, versus the initial star number $N$. We include a color bar to represent the average logarithm of the half-mass radius. In this figure, we compare the average values of different sets of simulations and the code used, including the direct $N$-body codes, shown as circle symbols in plot representing initial conditions from \citet{PortegiesZwartetal1999, Fujii2013, Katzetal2015, Wangetal2015, Mapelli2016, Sakuraietal2017, Reinosoetal2018, Panamarevetal2019, Reinoso2020, DiCarloetal2020a, Rastelloetal2021, Rizzutoetal2021a, Banerjee2021a, Gielesetal2021, Vergaraetal2021, Kamlahetal2022b, Rizzuto2023, Vergara2023, ArcaSeddaetal2023a, ArcaSeddaetal2024a, ArcaSeddaetal2024b, Vergara2025}, which are P+99, F+13, K+15, W+15, M+16, S+17, R+18, P+19, R+20, D+20, R+20, R+21, B+21, G+21, V+21, K+22, R+23, V+23, AS+23, and V+25, respectively. Simulations with Monte-Carlo codes are shown as square symbols in the plot representing initial conditions from \citet{Askaretal2017a, Rodriguezetal2019, Kremeretal2020a, Maliszewskietal2022, Rodriguezetal2022}, which are A+17, R+19, K+20, M+22, and R+22, respectively. Finally hybrid $N$-body simulations with diamond symbols in the plot represent the initial conditions from \citet{Wangetal2021, Wangetal2022a, Wang2024}, which are labeled as W+21, W+22, and W+24, respectively. The latter use the \textsc{PeTaR} code \citep{Wangetal2020a}, which is classified as a hybrid $N$-body code, because it combines direct particle-particle interactions with accelerated tree-based approximations. Particularly we include the following key benchmarks: the first \textit{N}-body simulation with one million body, known as the {\sc dragon} simulation, with an average density of $\rho_{h} \sim 10^4\,\rm M_\odot\,pc^{-3}$ \citep{Wangetal2015}; the {\sc dragon-II} simulations by \citet{ArcaSeddaetal2023a, ArcaSeddaetal2024a, ArcaSeddaetal2024b} exhibiting densities $\rho_{h} = 1.3\times 10^4$ to $6.9\times 10^5\,\rm M_\odot\,pc^{-3}$ for a particle range of $N = 1.2\times10^5-10^6$, with $10-33\%$ of primoridal binaries, and forming BHs with masses around $60-350\,\rm M_\odot$. Additionally, \citet{Vergara2023} simulated compact clusters with equal-mass stars, reaching densities up to $10^{10}\,\rm M_\odot\,pc^{-3}$ with fewer particles ranging from $N \sim 10^3 - 10^4$, forming massive objects of $60-68\,000\,\rm M_\odot$. The recent one-million-particle simulation by \citet{Vergara2025} achieved the highest recorded density to date $\rho_{h} = 6.9 \times 10^7\,\rm M_\odot\,pc^{-3}$, and produced an IMBH with a mass of $50\,000\,\rm M_\odot$. In the present work, we report simulations that reach even higher central densities, up to $\rho_{h} \sim 10^{10}\rm M_\odot\,pc^{-3}$.

\section{Results}

In this section, we analyze the cluster evolution for the different models, taking into account their varying numbers of particles and half-mass radii, and thus, their densities. We investigate their stellar dynamics, stellar evolution, and VMS/BH seed formation. 

\subsection{Dynamical evolution of stellar clusters}

In this subsection, we analyze the evolution of the cumulative mass that escapes from the star clusters, the number of collisions, and the Lagrangian radii at $90\%$, $50\%$, $30\%$, $10\%$, $5\%$ and $1\%$. 

The dynamics of stellar clusters are governed by key timescales,  the average collision timescale ($t_{coll}$) which determines the frequency of stellar collisions and is expressed as $t_{coll} = \sqrt{R / GM(n\Sigma_0)^2}$ \citep{Binney2008}, where $R$ and $M$ are the radius and the mass of the cluster, $G$ is the gravitational constant, $\Sigma_0$ is the effective cross-section and $n$ is the number density within the half mass radius; and the relaxation timescale ($t_{rx}$) which describes the time for the cluster to reach equilibrium via gravitational interactions and follows $t_{rx} = (0.1N / \ln{(\gamma N)}) t_{cross}$, where $t_{cross} = \sqrt{R^3 / GM}$, $\gamma$ is the Coulomb logarithm and $N$ the number of stars \citep{Binney2008}. From an analysis of observational data of NSCs, \citet{Escala2021} proposed that comparing relaxation and collision timescales ($t_{rx}$, $t_{coll}$) with the age of a cluster ($\tau$) determines whether collisions drive massive central object formation. \cite{Vergara2023} tested this using equal-mass star simulations, defining a critical mass when $t_{coll} = \tau$ to quantify the transition to collision-dominated evolution, leading to a critical mass $M_{crit} = R^{7/3} \left(4\pi M_* / 3\Sigma_0 \tau G^{1/2}\right)^{2/3}$, marking the onset of runaway stellar collisions when $M/M_{crit}\sim 0.1$.  

\begin{figure}[!h]
    \centering    
    \includegraphics[]{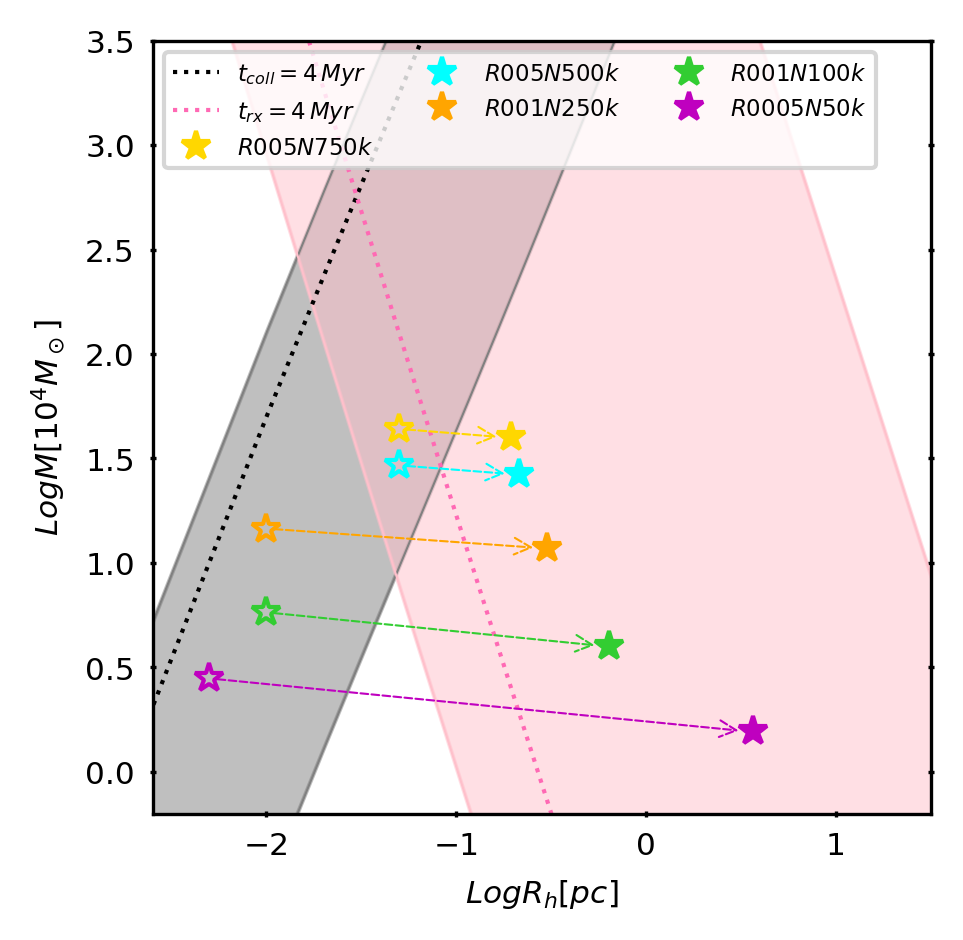}
    \caption{The dotted lines represent when the timescales ($t_{coll}$ and t$_{rx}$) are equal to $\tau = 4\,$Myr. The shaded regions indicate the parameter space where the respective timescales fall within the range $1\,$Myr $\leq \tau \leq 10\,$Gyr, highlighting different dynamical regimes: pink for two-body relaxation and grey for stellar collisions. Our clusters \texttt{R0005N50k}, \texttt{R001N100k}, \texttt{R001N250k}, \texttt{R005N500k}, and \texttt{R005N750k}, evolve for $3.73$, $3.81$, $3.64$, $3.80$, $3.87\,$Myr, respectively. We show their initial conditions with empty symbols and the final conditions with filled symbols, connected by arrows.}\label{fig_ic}
\end{figure}

In Fig.~\ref{fig_ic}, we present the dynamical regime of stellar systems in terms of their half-mass radius and total mass. We illustrate the interplay between both key dynamical timescales, $t_{coll}$ and $t_{relax}$ that fall within the range $1\,$Myr $\leq \tau \leq 10\,$Gyr, highlighting different dynamical regimes: pink for two-body relaxation and grey for stellar collisions. The lower limit of $\tau = 1\,$Myr corresponds approximately to the typical timescale for early cluster formation and the lifetimes of the most massive stars, while the upper limit of $\tau = 10\,$Gyr is set by the Hubble time, representing the maximum age over which stellar systems can dynamically evolve. The grey-shaded region is where collisions are significant throughout the lifetime of the cluster, they are practically from the very beginning in post-collapse evolution. It also illustrates the concept of a critical mass that increases steeply with radius, meaning that compact systems are more prone to entering the collisional regime. On the other hand, the pink-shaded region is where stellar systems \ff{are more likely to} evolve by relaxation and avoid collisions \citep{Escala2021, Vergara2023}. The overlapping region, dark pink, where both the two-body relaxation time and the stellar collision times fall between $1\,$Myr and $10\,$Gyr, suggests that systems in this region are neither extremely dense nor extremely diffuse, but instead evolve significantly through both two-body relaxation and direct stellar collisions over cosmic timescales. The shaded pink and black areas present an asymmetry that highlights that relaxation is a more widespread and efficient process across a wide range of stellar systems, whereas frequent stellar collisions require much more compact and dense conditions. 

We also include the initial half-mass radii and masses of our models (see Table~\ref{table1-ic}) connected with an arrow to the final half-mass radii and masses of the respective models \texttt{R005N750k}, \texttt{R005N500k}, \texttt{R001N250k}, \texttt{R001N100k} and \texttt{R0005N50k}, which are denoted by yellow, cyan, orange, green, and magenta star symbols, respectively (empty  for initial and full for final properties). Our models lie between the dotted lines (when $t_{coll}$ and $t_{rx}$ are equal to $\tau = 4\,$Myr). The models \texttt{R001N250k}, \texttt{R001N100k}, and \texttt{R0005N50k} fall within the collision-dominated regime. Their proximity to the dotted line indicates that they will experience several collisions, leading to the early formation ($< 4\,$Myr) of a massive central object \citep{Escala2021, Vergara2023}. In contrast, the \texttt{R005N750k} and \texttt{R005N500k} models lie in the dark pink region. These systems are also susceptible to collisions but will proceed through a more typical relaxation process. While the models within the grey-shaded area display highly stochastic behavior, those in the dark pink region evolve in a comparatively less chaotic manner.

The expansion of a star cluster is a well known phenomenon, driven primarily by internal dynamical processes but also by external tidal forces (see e.g. \citet{Fujii2016}). Over time, the escape rate from the cluster increases significantly as binary interactions become more frequent, indicating that stellar escapes are primarily driven by energy generated in these encounters. This gradual loss of stars, known as cluster evaporation, leads to a decrease in the mass of the cluster (see Appendix~\ref{Escapers_bin_and_hyp_Coll}). Additionally, binary stars can inject energy into the system through three-body encounters, further enhancing the expansion of the cluster. Our models evolve rightward and downward in parameter space as a result of expansion driven by mass loss through escapers. Despite being more compact due to computational limits, our simulated clusters host BH seeds ranging from a few thousand to tens of thousands of solar masses (see Fig.~\ref{fig:diff_bh_mass}). Observing stellar systems that host BHs remains challenging, as the high core brightness often obscures the possible presence of a black hole. Nonetheless, a BH candidate with a mass of at least $8200\,\rm M_\odot$ was recently identified at the center of $\omega$~Cen, based on observations of stars moving faster than the expected central escape velocity of the cluster \citep{Haberle2024}. 

\begin{figure}[!h]
    \centering
    \includegraphics[]{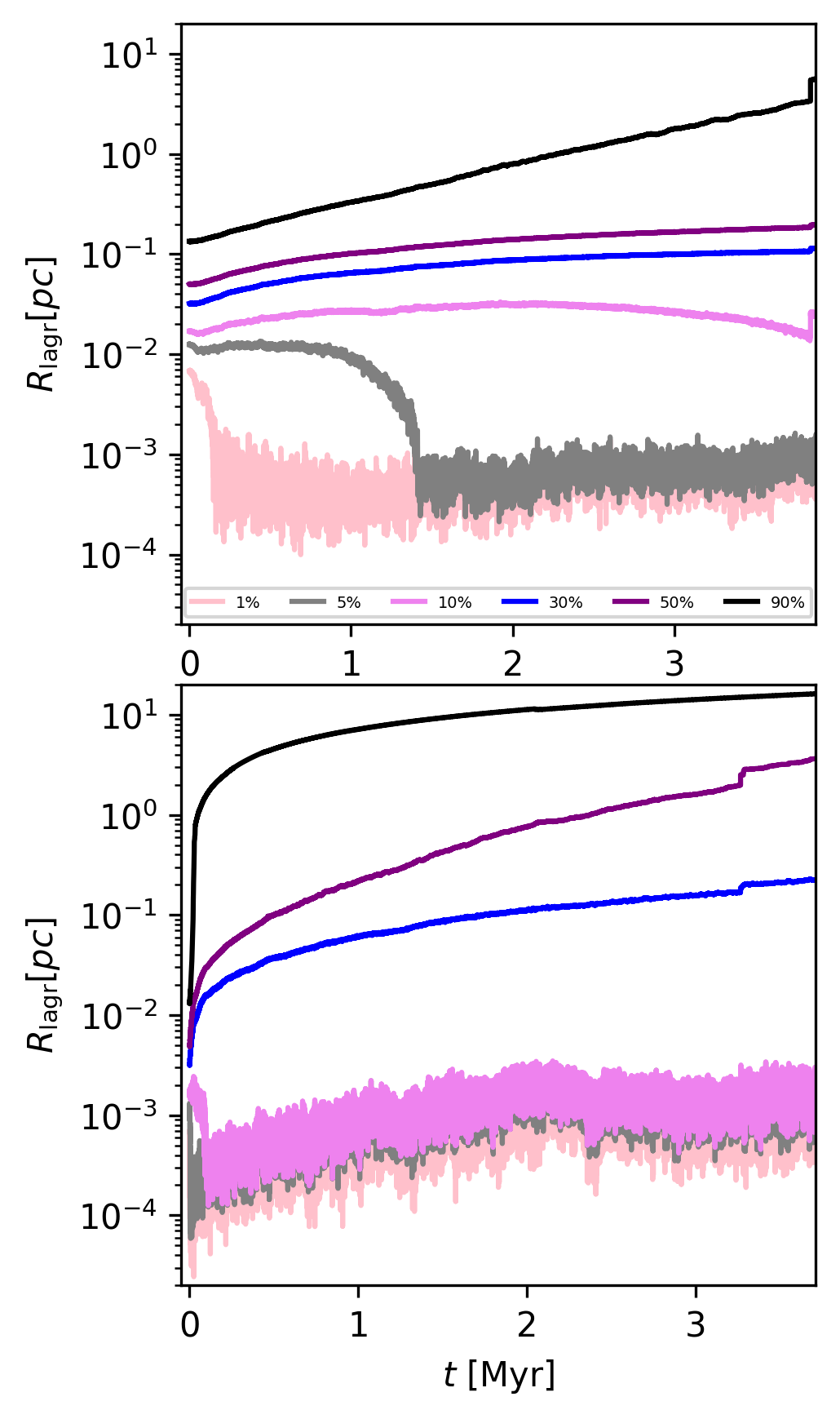} 
    \caption{Lagrangian radii calculated from the initial mass of the cluster for the $90\%$, $70\%$, $50\%$, $30\%$, $10\%$, and $1\%$ of the enclosed mass of model \texttt{R005N750k} (top) and model \texttt{R0005N50k} (bottom).}
    \label{R0005N50k-R005N750k}
\end{figure}

In Fig.~\ref{R0005N50k-R005N750k}, we display the evolution of two models, \texttt{R0005N50k} and \texttt{R005N750k}, the densest star cluster and the most massive star cluster, respectively. We present the temporal evolution of the Lagragian radii for models \texttt{R005N750k} and \texttt{R0005N50k}. The top panel shows the temporal evolution of model \texttt{R005N750k}; the Lagrangian radius at $1\%$ shows a sharp decline at the start of the simulation due to mass segregation, followed by a decrease of the $5\%$ after $1\,$Myr, while the $10\%$ shows a slight initial decrease before it evaporates smoothly and begins to decline slightly later. The $30\%$ and $50\%$ radii have an initial expansion which then continues steadily until the end. In contrast, the $90\%$ radius grows consistently throughout the simulation. The bottom panel shows that the $1\%$ and $5\%$ radii decrease sharply at the beginning of the simulation, while the $10\%$ begins its decline slightly later. Eventually, all three curves overlap, since the VMS controls the radii; the $30\%$ and $50\%$ radii have an initial more rapid expansion then smoothly continue expanding to the end, while the $90\%$ grows drastically at the beginning and then remains almost constant for the remainder of the simulation. Both models exhibit a steep initial decline in their innermost regions ($1\%$), although in \texttt{R0005N50k}, the $1\%$ and $5\%$ curves overlap at the start of the simulation; in \texttt{R005N750k}, the $5\%$ radius decreases more gradually after $1\,$Myr. In \texttt{R005N750k}, the $10\%$ curve expands smoothly before a late decline, while in \texttt{R0005N50k}, it drops off earlier and merges with the innermost curves. The $30\%$ and $50\%$ curves expand in both cases; however, in the \texttt{R0005N50k} model, it is faster than in \texttt{R005N750k}. Finally, \texttt{R0005N50k} shows a more abrupt initial growth in the $90\%$ curve, which then stabilizes, unlike the constant growth observed in \texttt{R005N750k}. These differences suggest that the initial conditions significantly influence the dynamical evolution of the system. Model \texttt{R005N750k} has a less violent behavior than model \texttt{R0005N50k}; this less violent process helps to retain more stars within the cluster, which is a reserve of stars that eventually can fall to the center, contributing to the growth of the VMS due to the constant stellar bombardment. On the other hand, \texttt{R0005N50k} presents a stochastic evolution, where several stars segregate almost intermediately to the center; however the low number of particles does not allow the VMS to be more massive.

\begin{figure}[!h]
    \centering
    \includegraphics[]{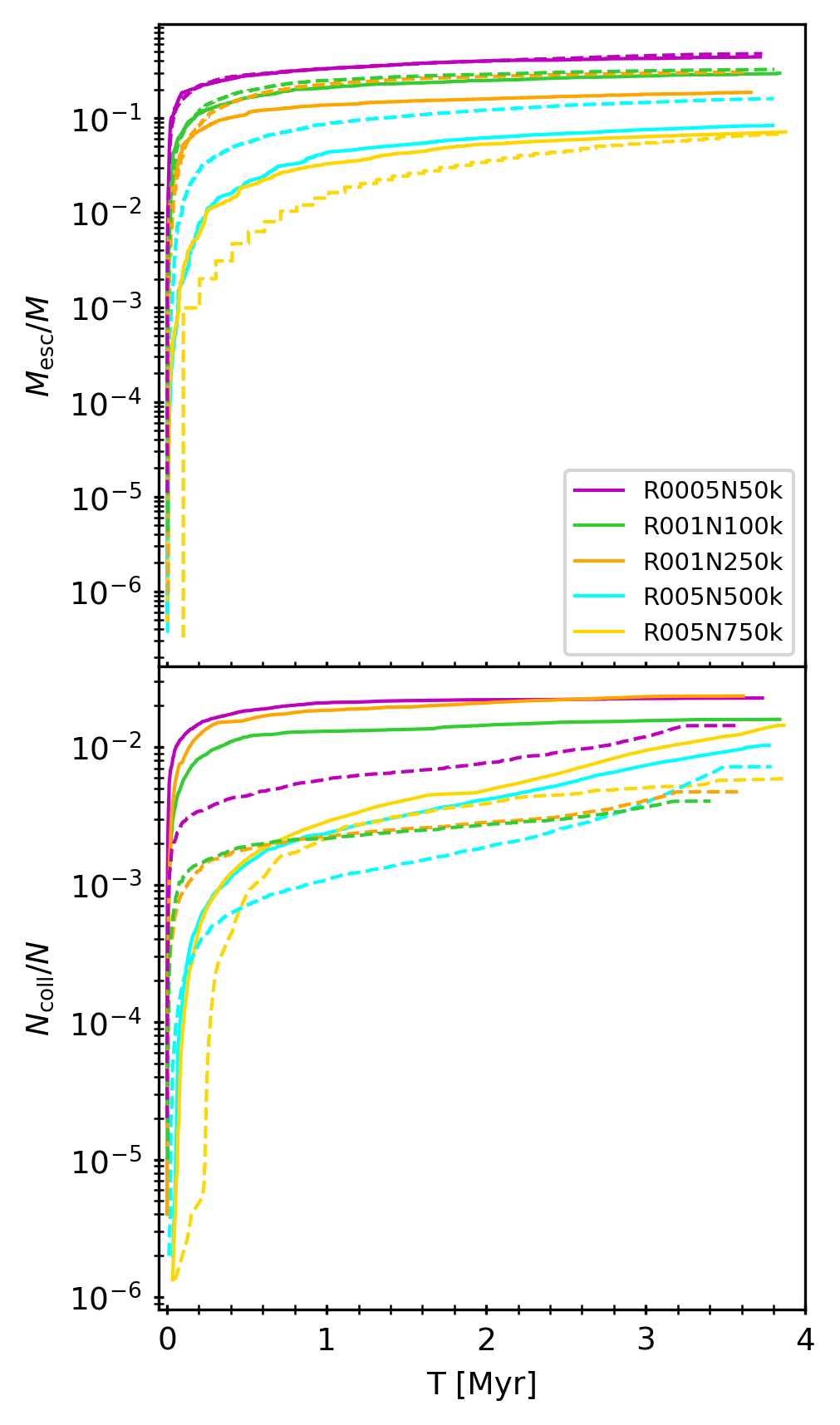} 
    \caption{Time evolution of the cumulative escaping mass normalized by the initial cluster mass (top) and the number of collisions with the most massive object} normalized by the initial particle numbers (bottom). Solid lines for \textit{N}-body and dashed lines for \textsc{MOCCA} simulations.
    \label{fig:diff_esc_n_coll}
\end{figure}

In Fig.~\ref{fig:diff_esc_n_coll}, we show the time evolution of the cumulative escaping mass normalized by the initial cluster mass (top panel) and the cumulative number of collisions involving the VMS, normalized by the initial number of stars (bottom panel) for our five cluster models; the solid lines represent the results of direct N-body simulations, while the dashed lines show the results of Monte Carlo simulations. 

In the top panel, we observe that clusters with fewer particles (e.g., model \texttt{R0005N50k}) exhibit higher escape mass fractions compared to those with more particles (e.g., model \texttt{R005N750k}). These lower-$N$ clusters are also denser, which leads to more dramatic dynamical evolution: the inner regions rapidly collapse at the beginning of the simulation, triggering multiple collisions. Additionally, enhanced three-body interactions increase the number of stars escaping from the system (see bottom panel in Fig.~\ref{R0005N50k-R005N750k}). In contrast, clusters with more particles (but lower density) show a lower ratio of cumulative escape mass normalized by the initial cluster mass, these systems exhibit a lower ratio of the number of binary collisions to the initial number of particles, thus the evaporation process is weaker (see Appendix~\ref{Escapers_bin_and_hyp_Coll}). The cumulative mass of escapers is $12\,371\,\rm M_\odot$, $17\,252\,\rm M_\odot$, $27\,173\,\rm M_\odot$, $24\,431\,\rm M_\odot$, and $31\,133\,\rm M_\odot$ in $N$-body simulations, and $13\,380\,\rm M_\odot$, $18\,888\,\rm M_\odot$, $44\,610,\rm M_\odot$\, $45\,640\,\rm M_\odot$, and $29\,984\,\rm M_\odot$ in Monte Carlo simulations for models \texttt{R0005N50k}, \texttt{R001N100k}, \texttt{R001N250k}, \texttt{R005N500k}, and \texttt{R005N750k}, respectively. 

In the bottom panel, models \texttt{R0005N50k},  \texttt{R001N100k} and \texttt{R001N250k}
with the smallest size ($R_{\mathrm{h}} \leq 0.01\,\mathrm{pc}$) and lowest particle number show the highest collision rate, consistent with its dense and rapidly collapsing core. Monte Carlo and \textit{N}-body results agree reasonably well, though the Monte Carlo simulations tend to slightly underestimate the number of collisions (see Appendix~\ref{Mocca_collisions_treatment} for details). Models \texttt{R005N500k} and \texttt{R005N750k}, representing the most massive and extended clusters, show the lowest per-star collision rates due to longer relaxation times, with similar trends in both simulation approaches. Denser clusters present higher ratios, however, since the collision counts are normalized by the initial number of stars, the ratio does not directly reflect the absolute number of collisions. The total number of collisions with VMS is $1135$, $1571$, $5956$, $5100$, $10\,919$ in \textit{N}-body and $716$, $404$, $1182$, $2859$, $4417$ in Monte Carlo simulations for models for models \texttt{R0005N50k}, \texttt{R001N100k}, \texttt{R001N250k}, \texttt{R005N500k}, and \texttt{R005N750k}, respectively.  Although the Monte Carlo method yields fewer total collisions, these typically involve more massive stars ($>100\,M_\odot$), while the \textit{N}-body simulations produce more frequent collisions involving lower-mass stars ($<1\,M_\odot$). For more discussion on this distinction, refer to Appendix~\ref{Collisions_and_mass}.

\subsection{Black hole formation efficiency}

In this subsection, we analyze the BH masses and their formation efficiency $\epsilon_{BH}=(1+M_f/M_{BH})^{-1}$, which is defined as the fraction of mass between the final mass of the cluster and the mass of the BH mass. This efficiency quantifies the fraction of the total stellar mass available in the system that eventually ends up in the BH, through stellar collisions, making it possible to explore the transition from a star-dominated cluster ($\epsilon_{BH}\sim0$) to a BH-dominated cluster ($\epsilon_{BH}\sim1$) as a function of the ratio of the initial mass to the critical mass $M/M_{crit}$.

\begin{figure}[!h]
    \centering
    \includegraphics[]{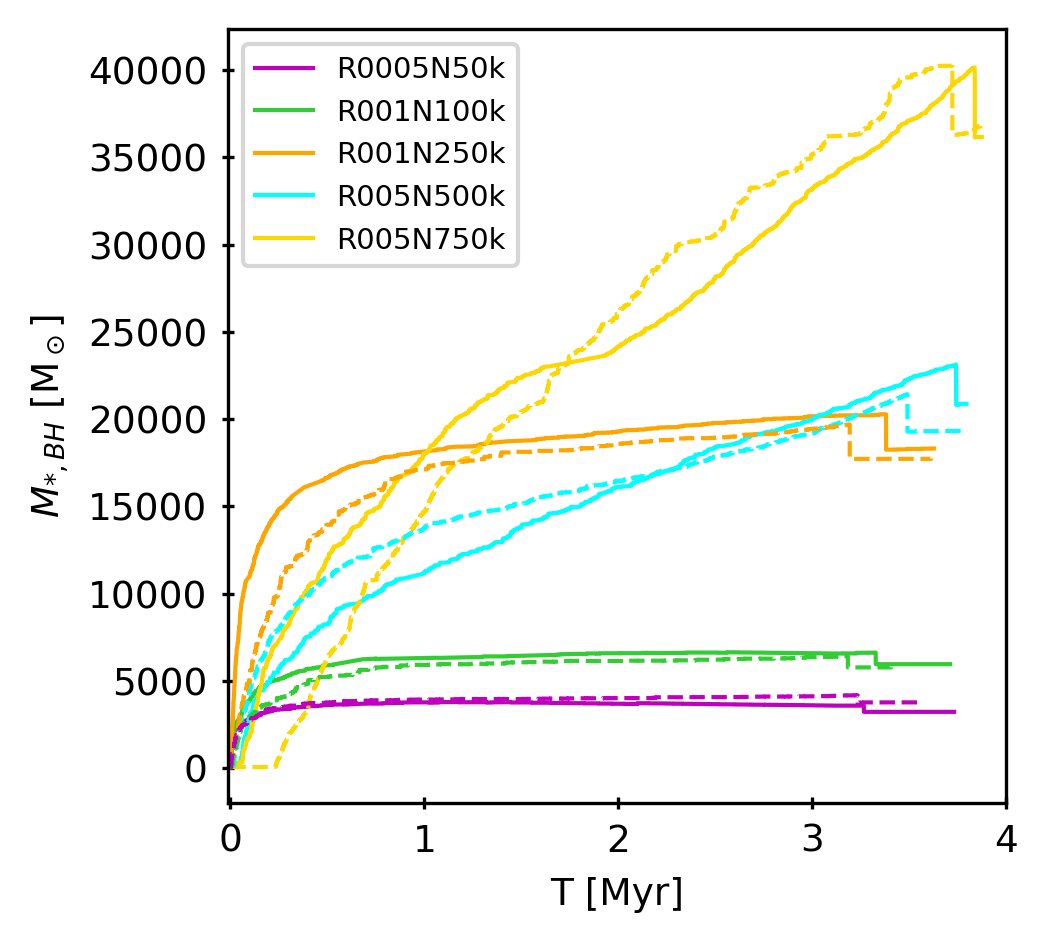} 
    \caption{Time evolution of the mass growth of VMSs that collapse into BH seeds. Solid lines for \textit{N}-body and dashed lines for \textsc{MOCCA} simulations. }\label{fig:diff_bh_mass}
\end{figure}

Fig.~\ref{fig:diff_bh_mass}, shows the mass evolution of the VMS/BH seeds, in our five cluster models. Solid lines indicate the results from direct N-body simulations, while dashed lines represent Monte Carlo simulations. In models \texttt{R005N750k} and \texttt{R005N500k},  both simulation types show a steep increase in mass with time, with Monte Carlo results reaching slightly higher values. Model \texttt{R001N250k}, shows an early rapid increase in mass, with a plateau around $~ 1 \,$Myr for both methods, presenting a slightly higher mass in \textit{N}-body simulation. Models \texttt{R001N100k} and \texttt{R0005N50k}, reach a much lower total mass, with their growth mostly flattening after $\lesssim 0.5\,$Myr. The VMS forms earlier in less massive clusters, whereas in more massive clusters the VMS is more massive, its growth is prolonged because of continued stellar bombardment. The formation of the BH seeds is delayed in clusters with larger particle numbers, as the larger stellar reservoir allows the VMS to rejuvenate more frequently, postponing its collapse. All curves show a characteristic drop when the BH seed is formed, corresponding to a loss of mass $10\%$ due to neutrino emission \citep{Fryeretal2012, Kamlahetal2022a}. We summarize the masses milestones for each model in Table~\ref{table2-VMS_formation}. 

\begin{table*}
\centering
\caption{The first column lists the model names, followed by columns showing the times ($\tau_{VMS,\,1k}$, $\tau_{VMS,\,5k}$,  $\tau_{VMS,\,10k}$,  $\tau_{VMS,\,20k}$, $\tau_{VMS,\,30k}$,  $\tau_{VMS,\,40k}$) at which the VMS reach certain mass milestones $\sim 1000, 5000, 10\,000, 20\,000, 30\,000, 40\,000\,\rm M_\odot$, respectively. \textsc{MOCCA} results are shown in round brackets next to \textit{N}-body results. A horizontal dash indicates that the VMS ended its life before reaching the corresponding mass milestone.}
\begin{tabular}{lcccccc}
\hline\hline
Models &  $\tau_{VMS,\,1k}$ &  $\tau_{VMS,\,5k}$ &  $\tau_{VMS,\,10k}$ &  $\tau_{VMS,\,20k}$ &  $\tau_{VMS,\,30k}$ &  $\tau_{VMS,\,40k}$ \\
&  [Myr] & [Myr] &  [Myr] &  [Myr]&  [Myr] &  [Myr] \\
\hline
\texttt{R005N750k}  & $0.069\,(0.27)$ & $0.16\,(0.45)$ & $0.38\,(0.68)$ &$1.16\,(1.44)$ & $2.69\,(2.31)$ & $3.81\,(3.58)$  \\
\texttt{R005N500k}  &  $0.063\,(0.03)$ & $0.20\,(0.11)$ & $0.74\,(0.41)$ &$3.01\,(3.18)$ & $-\,(-)$ & $-\,(-)$\\
\texttt{R001N250k}  & $0.0051\,(0.02)$ & $0.022\,(0.1)$ & $0.066\,(0.25)$ &$2.78\,(-)$  & $-\,(-)$ & $-\,(-)$ \\
\texttt{R001N100k}  & $0.0083\,(0.02)$ & $0.21\,(0.4)$ & $-\,(-)$ & $-\,(-)$& $-\,(-)$& $-\,(-)$   \\
\texttt{R0005N50k}  & $0.0031\,(0.01)$ & $-\,(-)$ & $-\,(-)$ & $-\,(-)$ & $-\,(-)$ & $-\,(-)$ \\
\hline\hline
\end{tabular}
\label{table2-VMS_formation}
\end{table*}

\begin{figure}[!h]
    \centering
    \includegraphics[]{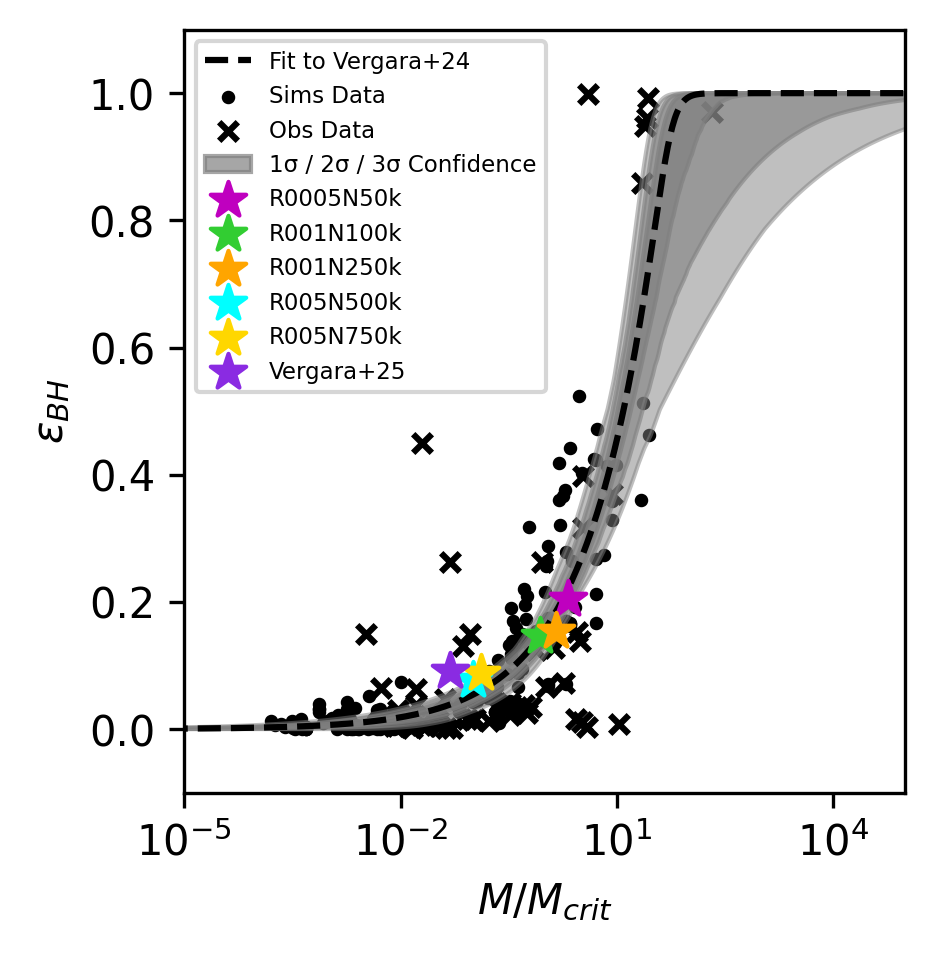} 
    \caption{BH formation efficiency as a function of the ratio of initial cluster mass to critical mass. The figure has been reproduced from \citet{Vergara2024}; we fit an asymmetric sigmoid function, represented by the black dashed line. The shaded grey bands represent the confidence intervals estimated by bootstrap resampling: the darker, middle, and lighter grey bands correspond to the $1\sigma$, $2\sigma$, and $3\sigma$ uncertainty regions. These intervals quantify the uncertainty in the fitted curve, arising from data variability. We add the simulation from \citet{Vergara2025} and the simulations from this work.}
    \label{fig:ebh_vs_Mcrit_adapted}
\end{figure}

In Fig. \ref{fig:ebh_vs_Mcrit_adapted}, we show the BH formation efficiency ($\epsilon_{BH}$) as a function of the ratio of the initial mass to the critical mass $M/M_{crit}$. The black dashed line represents an asymmetrical sigmoid function fitted to both numerical simulations and observational data collected in \citet{Vergara2024}, which includes diverse stellar systems, encompassing various initial conditions, initial mass functions, and evolution scenarios. To quantify the uncertainty in our nonlinear fit, we employ a bootstrap resampling method. By repeatedly resampling the original data with replacement $1000$ times and refitting the asymmetric sigmoid curve, we generate a set of fitted curves. From this set, we derive empirical confidence intervals at the uncertainties of the $1\sigma$, $2\sigma$, and $3\sigma$ regions, calculating the corresponding percentiles of the predicted curves at each point. These confidence bands provide a robust, nonparametric estimate of the uncertainty in the fitted model that naturally accounts for the noise and variability of the observed and simulated data collected in \citet{Vergara2024}. The form of the fit is given by,  $\epsilon_{BH}  = \left(1 + \exp\left[-k \left( \log\left(M/M_{crit}\right) - x_0 \right) \right]\right)^{a}$, where $k = 4.63$ controls the steepness of the transition, $x_0 = 4$ sets its location, and $a = -0.1$ determines the smoothness of the function.

The efficiency starts to increase when the initial mass approaches the critical mass, i.e., $M/M_{crit}\sim 0.1$ \citep{Vergara2024}. We include our models, those with higher densities exhibit the highest BH formation efficiency, even though their BH seeds are the lightest. Since $\epsilon_{BH}$ is a function between the BH mass and the remnant stellar mass. However, systems with more particles but lower densities show a lower efficiency. However, they experience more collisions, leading to heavier BH seeds compared to denser systems. We also include the simulation from \citet{Vergara2025}, which aligns well with the models presented in this work and the trend observed in \citet{Vergara2024}. The density plays an important role in the onset of collisions and VMS formation; however it is also limited by the number of stars available to sink to the center, so the formation of a VMS and subsequently a BH seed is influenced by these two parameters. In Table~\ref{table3-bh}, we summarize the BH seed masses, the formation time and the BH formation efficiency ($\epsilon_{BH}$).

\begin{table}
    \caption{The first column lists the model names, followed by the mass of the BH seed, then the time of BH formation, and finally the BH formation efficiency. \textsc{MOCCA} results are shown in round brackets next to \textit{N}-body results.}
    \begin{tabular}[{0.5\textwidth}]{lccc}\hline\hline
       Models & $M_{BH}$ & $\tau_{BH}$ & $\epsilon_{BH}$\\
       & $[10^3\mathrm{M_\odot}]$ & [Myr] & \\\hline
       \texttt{R005N750k}  & $36.1\,(36.2)$ & $3.83\,(3.72)$ & $8.87\,(8.87)\,\%$ \\
       \texttt{R005N500k}  & $20.8\,(19.2)$ & $3.74\,(3.48)$ & $7.74\,(7.80)\,\%$\\
       \texttt{R001N250k}  & $18.3\,(17.7)$ & $3.37\,(3.19)$ & $15.39\,(17.45)\,\%$ \\
       \texttt{R001N100k}  & $5.95\,(5.76)$ & $3.32\,(3.18)$ & $14.58\,(14.69)\,\%$\\
       \texttt{R0005N50k}  & $3.22\,(3.75)$ & $3.26\,(3.23)$ & $20.51\,(25.51)\,\%$\\
    \hline\hline    
    \end{tabular}
    \label{table3-bh}
\end{table}

\subsection{Comparison with JWST observations}

In this subsection, we compare our results with observational data including Young Star Clusters (YSCs) \citep{Brown2021}, as well as recent observations with JWST of YMCs \citep{Vanzella2022, Vanzella2022b, Vanzella2022c,Vanzella2023, Mowla2024, Adamo2024a}. 

\begin{figure*}
    \centering    
    \includegraphics[]{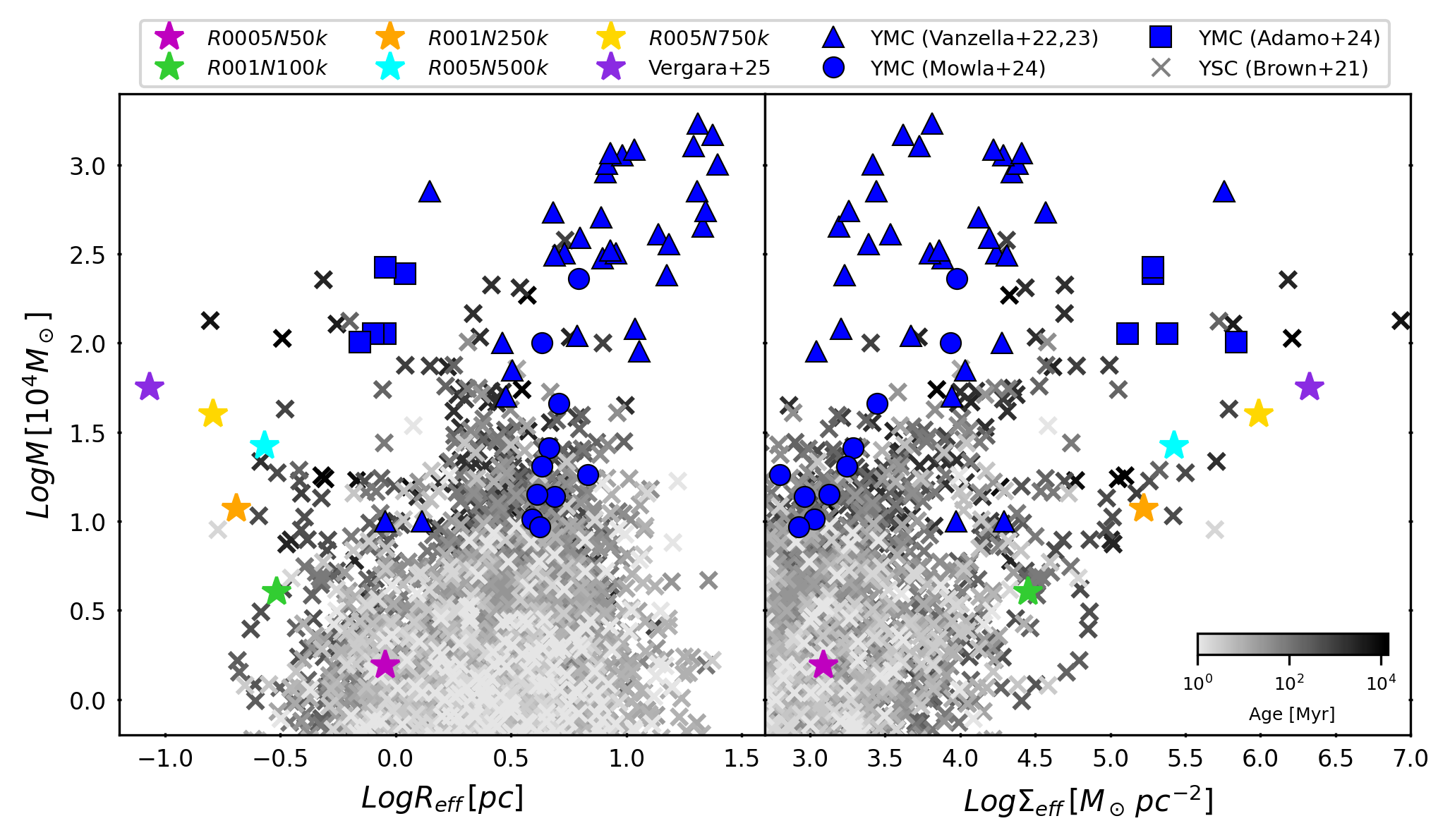}
     \caption{The final masses of our simulated clusters, along with the simulated cluster from \citet{Vergara2025}, are shown as star symbols. The current masses of YSCs from \citet{Brown2021} are also included, with the colorbar indicating their ages. Recent JWST observations of YMCs \citep{Vanzella2022, Vanzella2022b, Vanzella2023, Mowla2024, Adamo2024a} are shown with blue symbols, all plotted against the effective radius (left) and surface density (right).}\label{fig:obs}
\end{figure*}

On the left side of Fig.~\ref{fig:obs}, we show the effective radii and masses of YSCs which are denoted by grey crosses, while recent observations of YMCs with JWST are represented by blue symbols with black edges. YSCs span a wider range of the parameter space (see \citet{Brown2021}), but here we focus on those with masses between $10^4\,\rm M_\odot$ and $10^6\,\rm M_\odot$ and effective radii from $\sim 0.1\,\rm pc$ to $>10\,\rm pc$. YMCs have masses between $10^5$ and $10^8\,\rm M_\odot$ and effective radii from $<1$ to $>10\,\rm pc$. We compute the effective radii at the end of the simulations and also include our previous work from \cite{Vergara2025}. Our simulations (star symbols) fall into the mass range of $10^4$ and $10^6\,\rm M_\odot$, matching both the YSCs and the YMCs. The final effective radii range  between $\sim 0.1\,\rm pc$ and $1\,\rm pc$; our simulated cluster lies further to the left than the observations. However, our simulations are very short, around $4\,$Myr, so they do not have enough time to expand and thus move to the right in the plot. In particular, the varying shades of the grey crosses indicate their ages, with lighter colors representing younger ages and darker colors indicating older ones. The most massive YSCs typically exhibit smaller radii, whereas younger clusters tend to appear at the bottom, characterized by lower masses and larger radii (see \citet{Brown2021}, for details). This suggests that the most massive clusters expand more slowly, which is consistent with our simulations (see Fig.~\ref{fig_ic}). 

On the right side of Fig.~\ref{fig:obs}, we present the stellar surface densities and masses for our simulated clusters, alongside those from \citet{Vergara2025}. The simulated clusters span surface densities from approximately $10^3$–$10^6\,\rm M_\odot\, pc^{-2}$, similar to those observed in the local Universe, like YSCs, as well as YMCs at both low and high redshift, often regardless of their higher masses \citep[e.g.,][]{Adamo2024a}. In our simulations, the more massive clusters have both higher surface densities and slightly older ages, which is consistent with trends in the observed YSC populations \citep{Brown2021}. Observationally, a similar pattern emerges for YMCs; lower-mass clusters are typically younger (ages $\sim 3$-$4\,$Myr), as seen at $z\sim8$ in \citet{Mowla2024}, while the most compact and massive clusters are generally somewhat older, with ages up to about $36\,$Myr at $z\sim10$ \citep{Adamo2024a}. The YMCs studied by \citet{Vanzella2022, Vanzella2022b, Vanzella2023} at redshifts $z\sim2$-$6$ show a wide range of masses and ages, from $1$ to $30\,$Myr, further illustrating the diversity of cluster properties in the early Universe.

Overall, our simulations reproduce the observed trend that more massive and older clusters tend to have higher stellar surface densities, although the absolute values for YMCs can be substantially higher than those of lower-mass YSCs. This highlights the importance of both the cluster mass and evolutionary stage in shaping the structural properties of star clusters across cosmic time.
 

\subsection{Expected BH masses in YMCs detected via JWST}

Many of the observed YMCs have effective radii of a few $\rm pc$ and ages ranging from a few to several tens of Myr. Although they are massive stellar systems, it remains unclear whether their initial central densities were high enough to trigger efficient BH formation. While it is generally assumed that clusters were more compact at birth, the degree of this initial compactness remains uncertain. However, the critical mass exhibits a nonlinear dependence on cluster size and density: compact clusters reach this threshold at lower total masses. Thus, NSCs or YMCs can lie well near the $M_{\rm{crit}}$ region (see Fig.~\ref{fig_ic}), making them prime environments for collisional evolution and central object formation.

YMCs, though younger and potentially more transient than NSCs, can still reach extremely high densities shortly after formation \citep{Lahen2025}. If they are compact enough, the early phases of their evolution may fall within the regime where $t_{\mathrm{coll}} < t_{\mathrm{rx}} < \tau$. This opens a window during which massive stars can undergo collisions before they evolve off the main sequence, potentially forming exotic objects such as VMSs or heavy BH seeds. This makes $M_{\mathrm{crit}}$ a predictive tool not only for understanding cluster evolution, but also for identifying which systems might contribute to the formation of gravitational wave sources or seed BHs.

Based on the relationship between BH formation efficiency and the ratio of star cluster mass to critical mass, we estimate the expected minimum mass of massive BHs in these YMCs, under the assumption of a collision-driven formation scenario. In an initially more conservative estimate, we adopt the current mass and radius of the cluster to evaluate the initial to critical mass ratio, without accounting for potential early compactness or the presence of gas. The derived BH mass should therefore be considered a lower limit from the standpoint of stellar dynamics. Higher actual BH masses are not excluded, as additional growth could happen via accretion. In a more optimistic yet still realistic, scenario, we assume that the clusters  were initially more compact, as suggested by the scaling relation $r_h \propto t^{2/3}$ \citep{Fujii2016}\footnote{This relation holds for isolated clusters without mass loss \citep{GierszHeggie1996, GierszHeggie1997}. For tidally limited clusters, post core-collapse expansion can still be approximated by a power law $r_h \propto t^{(2-\mu)/3}$, where the index $\mu$ depends on the rate of mass loss.}. This would increase the stellar collision rate and thus the likelihood of forming more massive BHs. However, as noted above, it is not guaranteed that all of the observed clusters were compact enough to meet the conditions required by these models and specific properties of the clusters such as a higher binary fraction could also lead to a difference in their evolution. Furthermore, we neglect here the potential role of gas, which could enhance both the conditions for BH formation and subsequent BH growth through accretion.

\begin{figure}[!h]
    \centering
    \includegraphics[]{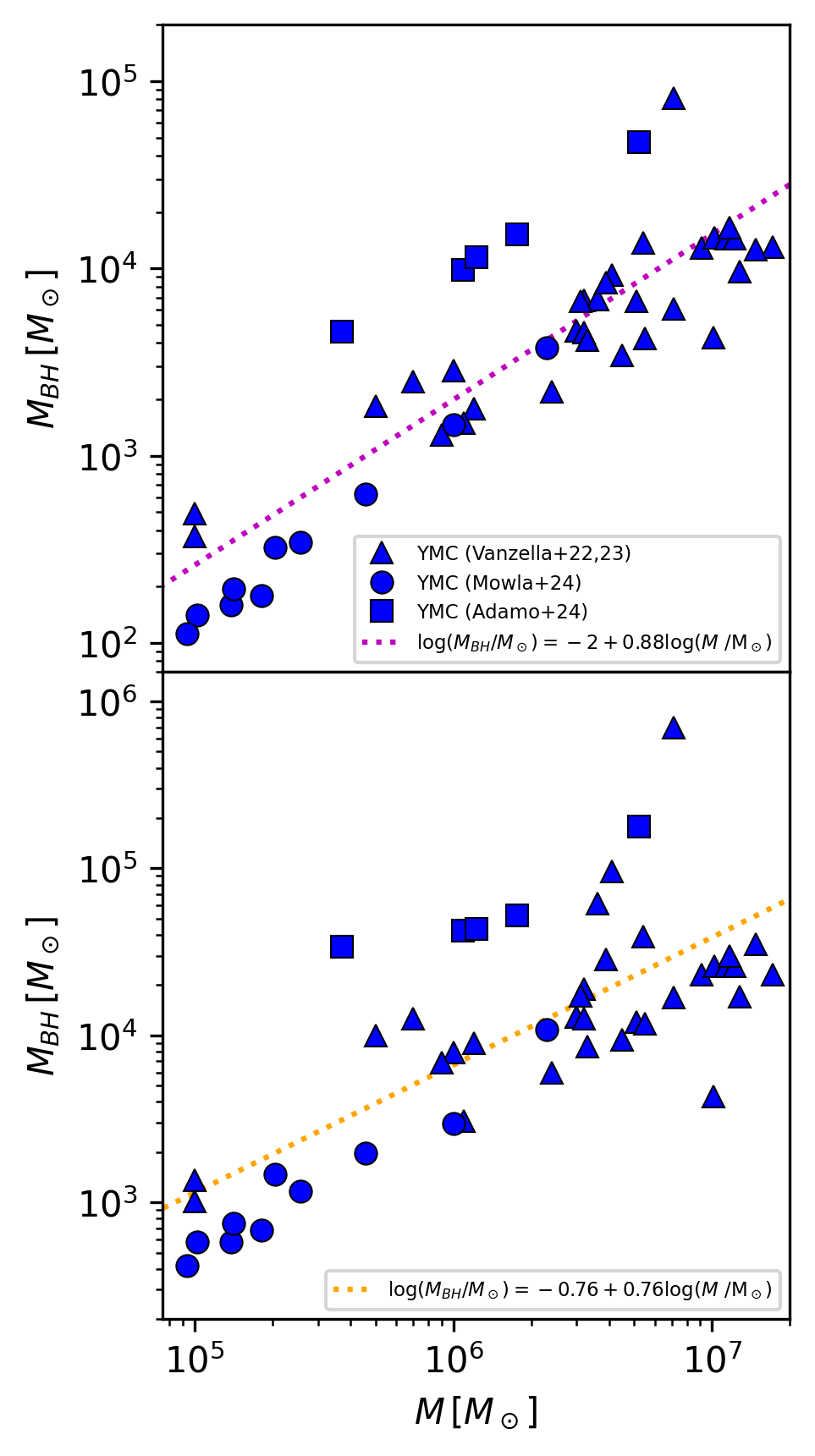} 
    \caption{Expected BH masses within YMCs against their current stellar mass. The top panel shows the conservative case, while the bottom panel shows the optimistic one.}
    \label{fig:exp_BH_YMCs}
\end{figure}

Fig.~\ref{fig:exp_BH_YMCs} shows the expected BH mass as a function of the stellar mass for the sources in the sample of \citet{Vanzella2022b, Vanzella2022, Vanzella2023, Adamo2024a, Mowla2024}. The more conservative prediction is shown on the top panel, while the more optimistic one is shown in the bottom panel. As the sample includes star clusters with different ratios of initial over critical masses, it includes both clusters where we expect high efficiencies of up to $1\%$, but also clusters with low efficiencies of order $0.01\%$. This is in the conservative case, while in the more optimistic case we expect efficiencies up to $10\%$. We find a relation between the mass of the cluster and the predicted mass of the BH that follows the shape of $\log(M_{BH}/M_\odot)=-2 + 0.88 \log(M~\rm/M_\odot)$, for the conservative case, and other with the shape of $\log(M_{BH}/M_\odot)=-0.76 + 0.76 \log(M~\rm/M_\odot)$, for the optimistic one. We expect that these predictions could be tested through a cross-correlation of the positions of the known sources with current and future X-ray data, which may lead to detections or at least upper limits for the BH masses that could be present within these sources. We also suggest that the presence of such a population of massive BHs at early times is likely to lead to relevant observables for the Laser Interferometer Space Antenna (LISA).

\section{Summary and conclusions}

We present simulations of the densest stellar clusters to date using \href{https://github.com/nbodyx/Nbody6ppGPU}{\textsc{Nbody6++GPU}} and the \href{https://moccacode.net/}{\textsc{MOCCA}} code. Our models include the latest updates for stellar radius evolution and close encounter prescriptions (see \citealt{Vergara2025}). We have also updated the stellar rejuvenation treatment to ensure that when a VMS collides with a less massive star, its age does not change dramatically (see appendix~\ref{rejuvenation} for details). Wind prescriptions follow \citet{Vinketal2001, VinkdeKoter2002, VinkdeKoter2005, Belczynskietal2010}, with the helium-burning phases modeled using luminosity-dependent winds \citep{Sanderetal2020}.

Our models represent stellar clusters with high density, and show BH seed formation through the purely collisional channel in a short period of time. Clusters with a higher initial number of stars experience more collisions, which can lead to the formation of more massive objects (e.g., models \texttt{R005N750k} and \texttt{R005N500k}). However, clusters with higher densities form massive objects more quickly (e.g., models \texttt{R0005N50k} and \texttt{R001N250k}). Clusters with more particles but lower densities undergo delayed core collapse and contain a larger reservoir of relatively massive stars that can migrate inward due to mass segregation. This results in more collisions over longer timescales, allowing the VMS to be repeatedly rejuvenated through mergers and postponing its collapse into a BH seed. In contrast, high-density clusters with fewer particles experience faster core collapse, triggering earlier collisions and rapid VMS growth. The VMS in these systems quickly increases its cross-section, further enhancing the collision rate. However, the stellar reservoir is depleted sooner, leading to the earlier collapse of the VMS into a BH seed.
Overall, our results show that the formation and evolution of VMSs (and their subsequent collapse into BH seeds) depend on both the cluster's density and its number of stars. Higher densities accelerate VMS formation, while a larger number of stars increases the eventual VMS mass. The timing of BH seed formation is regulated by the number of collisions that rejuvenate the VMS, thus, clusters with more stars form BH seeds later. Recent studies also indicate that mass loss in VMSs is highly sensitive to both the mass distribution of the colliding stars and the structural evolution of the VMS itself \citep{Ramirez-Galeano2025}.

Dense stellar clusters form in environments with low metallicity and high gas density \citep{Fukushima2023}, where radiative pressure is insufficient to prevent collapse. This leads to an extremely high star formation efficiency (SFE) of around $80\%$ \citep{Menon2023, Somerville2025}. Mergers of gas-rich dwarf galaxies at high redshift have also been proposed as a viable pathway for assembling massive, compact stellar systems \citep{renaud2015, lahen2020a, lahen2020b, Lahen2025}. Recent simulations suggest that, in such environments, the merger of star clusters hosting massive BHs can lead to the rapid formation of hard binary BH systems. These binaries interact strongly with surrounding stars and stellar-mass BHs, driving substantial ejection of both stellar and compact objects \citep{Souvaitzis2025}. Repeated ejections during and after cluster mergers can significantly reduce the final mass of the resulting cluster. These extremely dense stellar clusters are also ideal sites for the formation of supermassive stars through runaway collisions within their cores \citep{Charbonnel2023a}.

Our models exhibit stellar collision rates much higher than those typically found in globular cluster simulations, such as the {\sc dragon-II} runs \citep{ArcaSeddaetal2023a, ArcaSeddaetal2024a, ArcaSeddaetal2024b}. In our simulations, the timescale for collisions is shorter than the thermal timescale of the VMS, preventing the star from reaching thermal equilibrium. This can drive radial expansion, increase the star’s cross-section, and further enhance the collision rate. In addition, this process rejuvenates the VMS, delaying the formation of a BH seed. The thermodynamic state of such a VMS remains poorly understood. Mass loss could be driven by stellar winds  or by stellar mergers \citep{DaleDavis2006}, which also provide an additional enrichment pathway by ejecting processed material into the interstellar medium \citep{Gielesetal2018,Wang2020e}. However, the assumed rapid thermal recovery after each collision allows mass growth to outpace wind-driven mass loss. 

VMS formation through runaway stellar collisions in dense star clusters was first proposed analytically by \citet{Begelman1978, Rees1984}, and \citet{Lee1987}. Subsequent studies employed Fokker–Planck and Monte Carlo methods \citep{Quinlan1990, Gurkan2004, Freitag2006a, Freitag2006b, Gierszetal2015} and direct \textit{N}-body simulations \citep{Vergara2023, Vergara2025, Rantala2025, Rantala2025b}. In this study, dense stellar systems allow us to explore their dynamics and their ability to form VMSs in a short time due to the constant bombardment of stars sinking towards the center. The deep gravitational potential presented in these stellar systems leads to the onset of collisions that quickly trigger the formation of a VMS $\gtrsim 1000\,\rm M_\odot$ in less than $0.1\,$Myr. 

For the YMCs observed with JWST, the collision-based formation scenario implies BH masses in the range of $10^2$–$10^5,\rm M_\odot$ within these clusters. In our analysis, we considered both a conservative scenario (assuming the cluster masses and radii remain constant over time) and a more realistic scenario in which clusters undergo expansion, increasing the likelihood of forming massive objects due to higher initial densities. The numbers derived here are likely lower limits, as we did not account for the possible effects of gas. However, for more extended clusters, it is possible that massive object formation could be avoided in some cases. Moreover, the clusters observed with JWST are those that survived for a considerable period of time, while the densest clusters may have already collapsed and formed a massive BH at an earlier stage, as discussed by \citet{Escala2021}. The presence of these BHs could be confirmed or constrained with current and future X-ray observations. We also note that, while it is computationally challenging to model systems like the observed LRDs, their compactness naturally favors the collision-driven formation channel. It has been shown that the LRDs can have radii of $10\rm\,pc$ with core densities of the order of $10^8\rm\,M_\odot\,pc^{-3}$; note that this represents an upper and lower limit, respectively \citep{Guia2024}. Here we considered clusters with similar and higher central densities, but with smaller radii (and fewer particles), as otherwise due to computational limitations it is not feasible to explore these high densities.

TDEs, occurring when a star is torn apart by an IMBH’s tidal forces, produce luminous flares and provide insights into BH growth \citep{Stone2016, Ryu2020a, Ryu2020b, Ryu2020c, Stone2020}, which are particularly important for understanding LRD emissions \citep{Bellovary2025}.  Extreme mass ratio inspirals (EMRIs), in which stellar-mass BHs or neutron stars spiral into IMBHs via gravitational wave (GW) emission, are driven by dynamical friction and relativistic effects \citep{Hils1995, Broggi2022, Broggi2024}. Our models indicate rapid BH seed formation through runaway stellar collisions, suggesting EMRIs could occur earlier in cosmic history than predicted. Multi-messenger astronomy enables testing these theories: TDE flares are observed by the Hubble Space Telescope (\href{https://hubblesite.org/home}{HST}) \citep{Leloudas2016} and Neil Gehrels Swift Observatory (\href{https://swift.gsfc.nasa.gov}{Swift}) \citep{Brown2016}, while GW detectors (\href{https://www.ligo.caltech.edu/page/ligos-ifo}{LIGO}/ \href{https://www.virgo-gw.eu/}{Virgo}/ \href{https://gwcenter.icrr.u-tokyo.ac.jp/en/}{KAGRA}) \citep{Abbott2024} and future missions ((\href{https://lisa.nasa.gov}{LISA}) \citep{Amaro2017, McCaffrey2025}, and Einstein Telescope (\href{https://www.et-gw.eu}{ET}) \citep{Punturo2010}) target compact mergers and EMRIs.

Recent JWST observations of high-redshift galaxies have revealed unexpectedly high nitrogen abundances and elevated N/O ratios \citep{Tacchella2023a, NageleUmeda2023, Maiolino2024, Larson2023, Marques-Chaves2024, Naidu2025}, a phenomenon that standard chemical evolution models struggle to explain \citep{Cameron2023b}. While typical models predict low N/O ratios in young, low-metallicity systems (since nitrogen production is dominated by massive stars on longer timescales) the presence of VMSs offers a compelling solution. VMSs can rapidly enrich the interstellar medium with nitrogen through their strong winds and unique nucleosynthetic pathways, producing the high N/O ratios observed even in the early Universe \citep{Charbonnel2023a}. Incorporating VMSs into chemical evolution models allows both the amount and timing of nitrogen enrichment to better match observations, making them a key factor in understanding the chemical signatures of young, high-redshift stellar populations.
 
\begin{acknowledgements}

MCV acknowledges funding through ANID (Doctorado acuerdo bilateral DAAD/62210038) and DAAD (funding program number 57600326). MCV acknowledges the International Max Planck Research School for Astronomy and Cosmic Physics at the University of Heidelberg (IMPRS-HD).

AA acknowledges support for this paper from project No. 2021/43/P/ST9/03167 co-funded by the Polish National Science Center (NCN) and the European Union Framework Programme for Research and Innovation Horizon 2020 under the Marie Skłodowska-Curie grant agreement No. 945339. For the purpose of Open Access, the authors have applied for a CC-BY public copyright license to any Author Accepted Manuscript (AAM) version arising from this submission.

RS acknowledges NAOC International Cooperation Office for its support in 2023, 2024, and 2025, and the support by the National Science Foundation of China (NSFC) under grant No. 12473017. This research was supported in part by grant NSF PHY-2309135 to the Kavli Institute for Theoretical Physics (KITP).

FFD and RS acknowledge support by the German Science Foundation (DFG, project Sp 345/24-1).

MAS acknowledges funding from the European Union’s Horizon 2020 research and innovation programme under the Marie Skłodowska-Curie grant agreement No.~101025436 (project GRACE-BH, PI: Manuel Arca Sedda). MAS acknowledge financial support from the MERAC foundation.

DRGS gratefully acknowledges support by the ANID BASAL project FB21003 and ANID QUIMAL220002. DRGS thanks for funding via the  Alexander von Humboldt - Foundation, Bonn, Germany.

AE acknowledges financial support from  the Center for Astrophysics and Associated Technologies CATA (FB210003).

MG was supported by the Polish National
Science Center (NCN) through the grant 2021/41/B/ST9/01191.

Computations were performed on the HPC system Raven at the Max Planck Computing and Data Facility, and we also acknowledge the Gauss Centre for Supercomputing e.V. for computing time through the John von Neumann Institute for Computing (NIC) on the GCS Supercomputer JUWELS Booster at Jülich Supercomputing Centre (JSC).

We also acknowledge A. Sander and his team for helpful comments.
\end{acknowledgements}

\section*{Data Availability}
The underlying data, including the initial model used in this work, as well as the output and diagnostic files from both the \textsc{Nbody6++GPU} and \textsc{MOCCA} simulations, will be shared on reasonable request. 



%
\bibliographystyle{aa} 
\bibliography{ref} 
%
\begin{appendix}

\section{Treatment of the rejuvenation of a star through collisions} \label{rejuvenation}
   
The rejuvenation of a main-sequence star upon collisions with other main-sequence stars was numerically implemented by \citet{Hurleyetal2002a, Hurleyetal2005}. The main motivation for the introduction was to provide a proper treatment of blue straggler stars. These stars consistently find themselves above a stellar populations MS turn-off point \citep[see e.g., for observational evidence, check][]{Giesersetal2018}. This implies that the effective age, $A_{\mathrm{MS}}$, of these MS stars after a collision and a merger, is younger than the average age of the stellar population. In our models, which provide many collisions of a VMS with small mass MS stars, this treatment is not correct for the VMS, as the mass ratio $q$,
\begin{equation}
\label{eq:massratio}
q = \frac{M_{\mathrm{MS,2}}}{M_{\mathrm{MS,1}}} \quad 
\end{equation}
becomes very small (i.e., $q\ll 1$). 

If we work under the base assumption that hydrogen in the MS core is uniformly distributed, then,  after a collision, we have a convective mixing of the hydrogen fuel from envelope to core. The \textsc{BSE} code terminates the MS phase of the star and enters the Hertzsprung gap (HG) phase after $10~\%$ of the total amount of hydrogen in the MS star has been burned \citep{Hurleyetal2000, Hurleyetal2005}, the computational application is described below.
We first consider the collision of two MS stars with (effective) ages $A_{\mathrm{MS,1}}$ and $A_{\mathrm{MS,2}}$, MS life-times $\tau_{\mathrm{MS,1}}$ and $\tau_{\mathrm{MS,2}}$, as well as masses $M_{\mathrm{MS,1}}$ and $M_{\mathrm{MS,2}}$, respectively. Here, the subscripts $1$ and $2$ denote the primary (and more massive) MS star and the secondary star, respectively. The collision produces the "third" MS star with effective age $A_{\mathrm{MS,3}}$, MS life time $\tau_{\mathrm{MS,3}}$ and mass $M_{\mathrm{MS,3}}$. If we know these parameters, \citep[as suggested in][]{Hurleyetal2005} we can evaluate the final effective age $A_{\mathrm{H,MS,3}}$ of the produced MS star with their equation \citep[see also][for more information]{Glebbeeketal2008}.This treatment works well for MS collisions that have mass ratios $q \approx 1$, which might be expected for many blue stragglers. \\ 

Here we have updated the traditional treatment, finding the age of the newly formed MS star as:

\begin{eqnarray} 
\label{eq: rejuvenation 1}
    F_q &=& 1.0 - q\cdot(1.0-0.1/(1+q)),\\
\label{eq: rejuvenation 2}
    A_{\mathrm{MS,3}} &=& \tau_{MS,3} \cdot F_q \cdot (A_{\mathrm{MS,1}}/\tau_{MS,1} - q\cdot A_{\mathrm{MS,2}}/\tau_{MS,2}).
\end{eqnarray}
This treatment for the age of the VMS ensures that (i) if $q\rightarrow 1$ and $F_q \rightarrow 1$ (that is, the two colliding stars have similar masses), we get Eq.~\ref{eq: rejuvenation 2}, while (ii) for $q\rightarrow 0$ and $F_q \rightarrow 0$  (that is, the primary star is much more massive than the secondary star) the age of the VMS does not change significantly. This approach improves on the traditional treatment by ensuring a physically reasonable behavior in the limit $q\ll 1$. In such cases, the secondary star contributes very little mass, and thus the merger product should retain an age close to that of the more massive primary. The introduction of the factor $F_q$ ensures that the rejuvenation is suppressed when $q$ is small, preventing an unrealistically young apparent age for the merger product. In contrast, earlier models lacking such a correction could significantly underestimate the age in this regime, resulting in overly rejuvenated stars even when the secondary's contribution is negligible. 

\section{Treatment of collisions in \textsc{MOCCA}} \label{Mocca_collisions_treatment}

\begin{figure}[!h]
    \centering
    \includegraphics[width=0.34\textwidth]{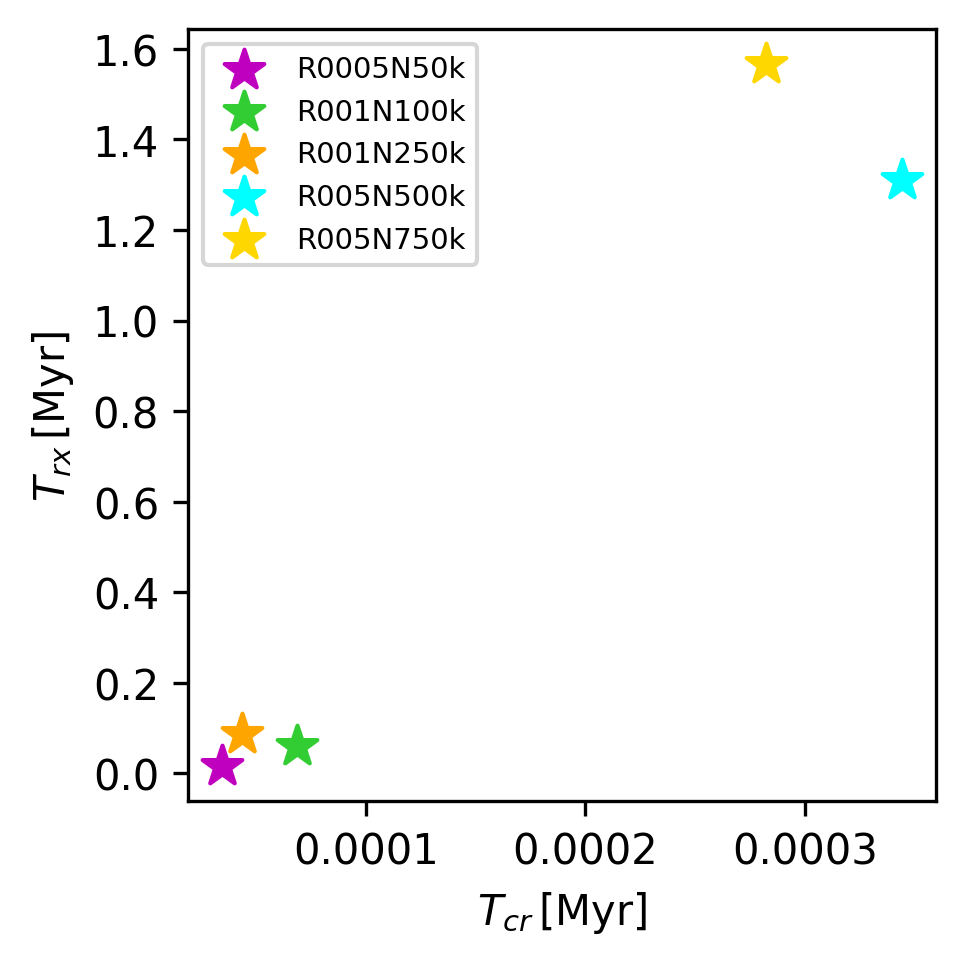}
    \caption{
        Initial half-mass relaxation time ($T_{\mathrm{rx}}$) versus initial crossing time ($T_{\mathrm{cr}}$) for the five cluster models simulated in this work. The three models with the shortest relaxation and crossing times (\texttt{R0005N50k}, \texttt{R001N100k}, and \texttt{R001N250k}) required a modified collision treatment, since the standard local-density-based prescription significantly overestimates the collision rate in these extreme regimes
    }
    \label{fig:cross_relax_times}
\end{figure}

In the models with the shortest crossing and relaxation times considered in this work, specifically the \texttt{R0005N50K}, \texttt{R001N100K}, and \texttt{R001N250K} runs (see Fig.\,\ref{fig:cross_relax_times}), we found that the standard MOCCA collision treatment, based on local density and cross-section estimates, leads to unrealistically rapid growth of the VMS. In particular, before our improvements, the mass of the VMS would increase by more than a factor of two to three, compared to results from direct \textit{N}-body simulations, indicating that collisions with VMS were overpredicted in these extreme regimes. To address this, we developed and implemented an improved, physically motivated treatment for VMS collisions in these high-density, low-\textit{N} models. After introducing this new prescription, the MOCCA results for VMS growth were brought into good agreement with direct \textit{N}-body simulations.

In standard MOCCA simulations, the probability for a collision between two single stars is computed using the cross-section formalism:
\begin{equation}
    P_{\mathrm{coll}} = 0.5\,C\,p_{\mathrm{max}}^2\,n\,u\,\Delta t
\end{equation}
where $C$ is a normalization constant, $p_{\mathrm{max}}$ is the maximum impact parameter (related to the sum of stellar radii and gravitational focusing), $n$ is the local number density, $u$ is the relative velocity, and $\Delta t$ is the timestep. The collision outcome is then determined probabilistically, and, if a collision occurs, the two stars are merged.

For these compact models ($R_{\mathrm{h}} \leq 0.01\,\mathrm{pc}$) with relatively low number of stars, this cross-section-based approach becomes inaccurate because the presence of a VMS can dominate the local gravitational potential, invalidating the assumptions of the local density formalism and leading to overestimated collision rates. To address this limitation, we introduced a more physically motivated, two-body treatment for collisions involving VMSs, defined here as stars with $M \geq M_{\mathrm{VMS,th}}$, where $M_{\mathrm{VMS,th}} = 500\,\rm M_\odot$ in our production runs. This new approach assumes that, in the presence of a dominant central VMS, gravitational focusing and two-body orbital dynamics, rather than the local stellar density, govern whether a close passage results in a physical collision. In this approach, instead of applying the probabilistic cross-section formula, we explicitly compute the Keplerian orbital parameters for stars that are closest to each VMS in the system.

For each collision candidate pair where the primary star satisfies $M_1 \geq M_{\mathrm{VMS,th}}$, we compute the specific energy and specific angular momentum of the secondary star (hereafter, ``star 2'') with respect to the VMS (``star 1''):
\begin{align}
    \epsilon_2 &= \frac{1}{2} \left(v_{r,2}^2 + v_{t,2}^2 \right) - \frac{G M_1}{r_2} \\
    \ell_2 &= r_2\,|v_{t,2}|
\end{align}
where $v_{r,2}$ and $v_{t,2}$ are the radial and tangential velocities of star 2 with respect to the VMS, $r_2$ is its distance from the VMS, and $G$ is the gravitational constant (set to unity in code units).

The pericenter distance of star 2 with respect to the VMS is then given by
\begin{equation}
    r_{\mathrm{p},2} = \frac{ -G M_1 + \sqrt{ (G M_1)^2 + 2\epsilon_2 \ell_2^2 } }{ 2\epsilon_2 }.
\end{equation}
If $\epsilon_2 \approx 0$ (parabolic approach), we instead use
\begin{equation}
    r_{\mathrm{p},2} = \frac{ \ell_2^2 }{ 2 G M_1 }.
\end{equation}
A physical collision is deemed to occur if $r_{\mathrm{p},2} \leq R_1 + R_2$, where $R_1$ and $R_2$ are the radii of the two stars. In this case, the collision is forced ($P_{\mathrm{coll}} = 1$), bypassing the original probabilistic cross-section calculation. If the pericenter is greater than the sum of the radii, no collision occurs.

The modified treatment is applied only when the more massive star in the pair exceeds the VMS mass threshold. For other star pairs, we use the original cross-section-based approach. For efficiency, and to mimic the high rate of encounters expected in extreme environments, this pericenter criterion is applied to a fixed number of the nearest neighbours of the VMS during each timestep. Specifically, we use the 30 nearest stars for the \texttt{R0005N50k} and \texttt{R001N100k} models, and 100 neighbours for the \texttt{R001N250k} run. All other aspects of the collision and merger handling, such as mass loss, rejuvenation, and product type, remain as in the standard MOCCA prescription.

We also applied this modified routine to the \texttt{R005N500k} model, where the pericenter test was used for the 200 nearest neighbours of each VMS. In that case, the combination of a larger neighbour set and a smaller timestep led to significantly improved agreement between MOCCA and direct $N$-body results for VMS growth. Without the new treatment, for the \texttt{R005N500k} run, the standard collision prescription overestimated the IMBH mass at formation by about $5{,}000,M_\odot$ ($25{,}000,M_\odot$ versus $20{,}000,M_\odot$ in the direct $N$-body simulation). 

For the \texttt{R0075N750k} model, however, we found that the new prescription underpredicted the final IMBH mass relative to the direct $N$-body result. Therefore, for this highest-$N$ case, we retained the original MOCCA collision treatment. As shown previously by \citet{Vergara2025}, the agreement between MOCCA and direct $N$-body simulations is reasonable for large-$N$ models when the standard probability-based collision treatment is used. In such high-$N$ clusters, the larger number of massive stars provides a significant reservoir of potential collision partners for the VMS, enabling continuous growth over several Myr. Additionally, the longer relaxation time in these systems means that mass segregation and the subsequent delivery of massive stars to the cluster center proceeds more gradually, supporting a more extended phase of VMS growth.

In summary, we replaced the cross-section-based probabilistic collision criterion with a two-body pericenter test for VMSs, as described by the equations above. This suppresses artificially high collision rates in extreme-density, low-$N$ runs and yields a more physical growth rate for VMSs, bringing the MOCCA results into agreement with the direct $N$-body simulations. We emphasize that this represents only a first step toward developing an improved and more general treatment for collisions involving a VMS or a central IMBH. In future work, we plan to implement and test more sophisticated prescriptions, guided by systematic comparisons with direct $N$-body simulations.

\section{Number of collisions and mass contribution} \label{Collisions_and_mass}

We analyze the number of stellar collisions and their mass contributions to VMS and BH seed formation, from both simulation codes.

\begin{figure}[!h]
    \centering
    \includegraphics[]{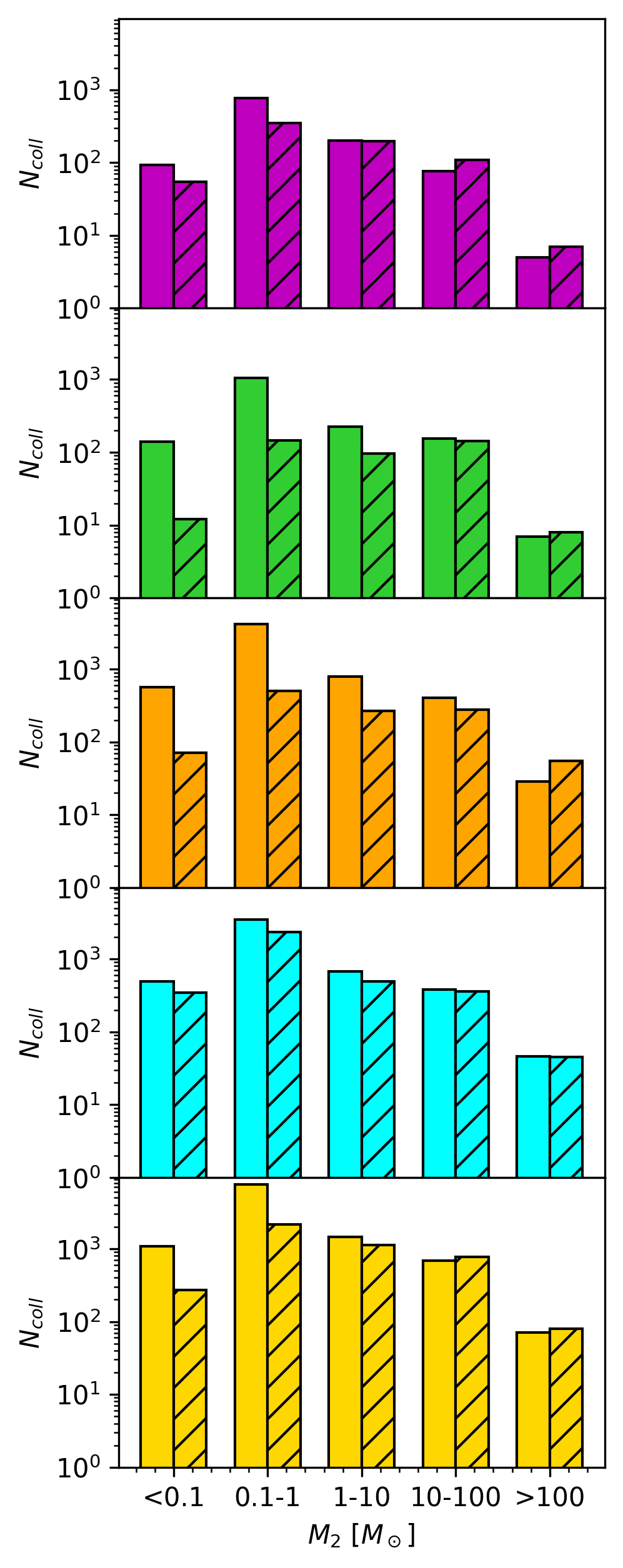} 
    \caption{Histograms showing the number of collisions contributing to VMS/BH seed formation. The mass bins are: $<0.1~\mathrm{M}_\odot$, $0.1$–$1~\mathrm{M}_\odot$, $1$–$10~\mathrm{M}_\odot$, $10$–$100~\mathrm{M}_\odot$, and $>100~\mathrm{M}_\odot$. Results are shown for both the N-body simulations (solid bars) and the MOCCA simulations (hatched bars). From top to bottom, the models are \texttt{R0005N50k}, \texttt{R001N100k}, \texttt{R001N250k}, \texttt{R005N500k}, and \texttt{R005N750k}.}
    \label{fig:hist_coll}
\end{figure}

Fig.~\ref{fig:hist_coll} presents histograms of the number of collisions across different mass ranges. The results from the \textit{N}-body simulations are shown with solid bars, while those from the \textsc{MOCCA} code are indicated with hatched bars. The histogram bins correspond to the following mass intervals: $<0.1$, $0.1$–$1$, $1$–$10$, $10$–$100$, and $>100~\mathrm{M_\odot}$. Each panel represents a different model from top to bottom: \texttt{R0005N50k}, \texttt{R001N100k}, \texttt{R001N250k}, \texttt{R005N500k} and \texttt{R005N750k}. The number of collisions for the different mass range in each models is listed below, the first set corresponds to the \textit{N}-body, and the second to the \textsc{MOCCA} results:

\begin{itemize}
    \item Model \texttt{R0005N50k}:\\
    \textit{N}-body: 92, 762, 200, 76, 5\\
    \textsc{MOCCA}: 54, 351, 193, 108, 7
    \item Model \texttt{R001N100k}:\\
    \textit{N}-body: 139, 1046, 223, 156, 7\\
    \textsc{MOCCA}: 12, 145, 96, 143, 8
    \item Model \texttt{R001N250k}:\\
    \textit{N}-body: 571, 4156, 796, 404, 29\\
    \textsc{MOCCA}: 71, 508, 268, 280, 55
    \item Model \texttt{R005N500k}:\\
    \textit{N}-body: 491, 3503, 677, 383, 46\\
    \textsc{MOCCA}: 342, 2357, 494, 358, 45
    \item Model \texttt{R005N750k}:\\
    \textit{N}-body: 1072, 7649, 1442, 685, 71\\
    \textsc{MOCCA}: 271, 2176, 1124, 767, 79
\end{itemize}

\begin{figure}[!h]
    \centering
    \includegraphics[]{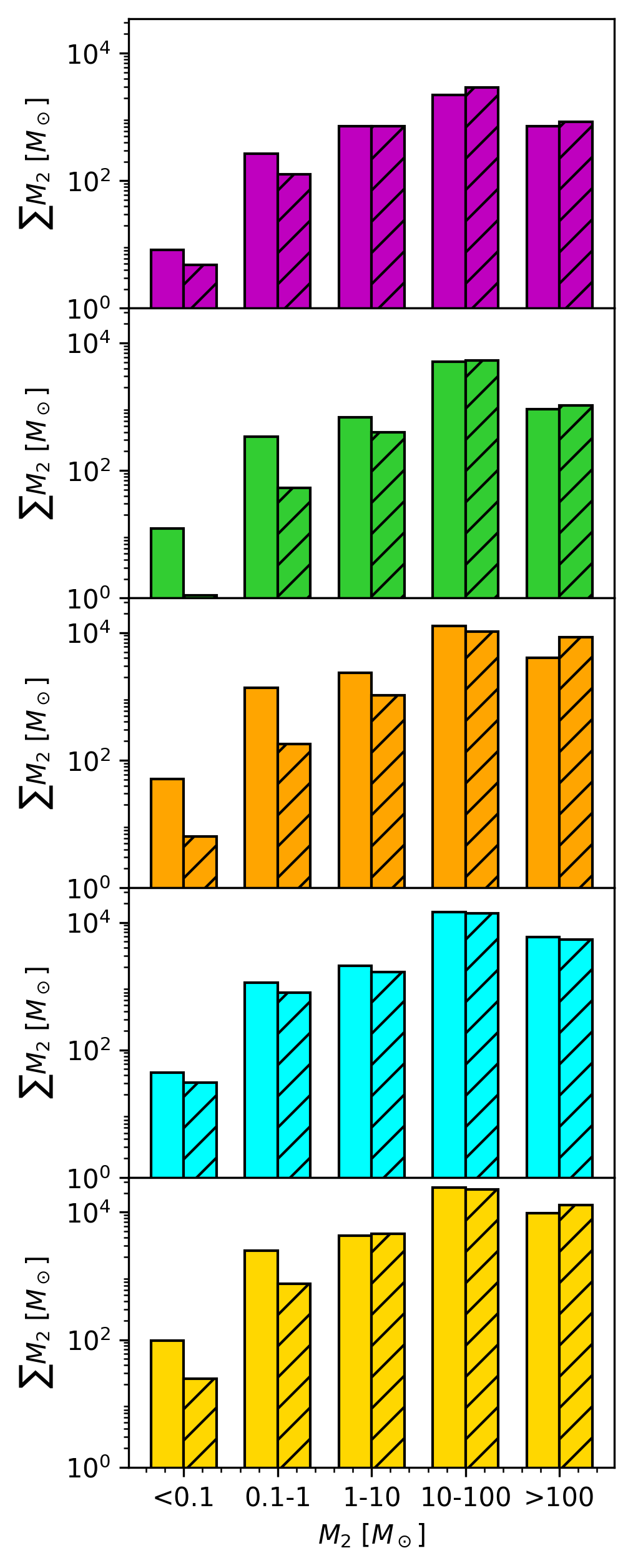} 
    \caption{Histograms showing the cumulative mass contributed to VMS/BH seed formation. The mass bins and bar styles are the same as in Fig.~\ref{fig:hist_coll}.}
    \label{fig:hist_mass}
\end{figure}

Fig.~\ref{fig:hist_mass} presents histograms of the mass contribution to the VMS and BH seed formation. The bar styles, panels, bin mass intervals, and bullet points are the same as in Fig.~\ref{fig:hist_coll}, the results are:

\begin{itemize}
    \item Model \texttt{R0005N50k}:\\
    \textit{N}-body: 8.20, 265.42, 722.30, 2237.57, 709.96$~\mathrm{M_\odot}$ \\
    \textsc{MOCCA}: 4.74, 125.72, 715.84, 2921.03, 842.85$~\mathrm{M_\odot}$
    \item Model \texttt{R001N100k}:\\
    \textit{N}-body: 12.41, 336.44, 682.50, 5082.01, 920.88$~\mathrm{M_\odot}$\\
    \textsc{MOCCA}: 1.09, 53.68, 396.52, 5344.29, 1055.53$~\mathrm{M_\odot}$
    \item Model \texttt{R001N250k}:\\
    \textit{N}-body: 50.99, 1377.53, 2355.33, $12\,805.41$, 4039.95$~\mathrm{M_\odot}$\\
    \textsc{MOCCA}: 6.32, 180.35, 1042.07, $10\,343.75$, $8502.78~\mathrm{M_\odot}$
    \item Model \texttt{R005N500k}:\\
    \textit{N}-body: 43.77, 1146.33, 2067.01, $14\,622.21$, 5835.47 $~\mathrm{M_\odot}$\\
    \textsc{MOCCA}: 30.56, 789.12, 1671.15, $13\,943.06$, 5420.96$~\mathrm{M_\odot}$
    \item Model \texttt{R005N750k}:\\
    \textit{N}-body: 95.80, 2490.73, 4241.25, $24\,527.13$, 9636.05$~\mathrm{M_\odot}$\\
    \textsc{MOCCA}: 24.21, 754.99, 4565.79, $22\,993.53$, $12\,946.69~\mathrm{M_\odot}$
\end{itemize}

Both codes produce different numbers of collisions within the mass interval bins, resulting in varying mass contributions, despite starting from the same initial conditions. The models \texttt{R0005N50k}, \texttt{R001N100k}, \texttt{R001N250k}, \texttt{R005N500k}, and \texttt{R005N750k} initially contain 5, 5, 30, 57, and 79 stars with masses $>100~\mathrm{M}\odot$, respectively. However, in the \textsc{MOCCA} simulations, the VMS shows a greater contribution from stars with masses $>100~\mathrm{M}\odot$ in the \texttt{R0005N50k}, \texttt{R001N100k}, and \texttt{R001N250k} models. This occurs because the standard \textsc{MOCCA} prescription forms massive stars earlier, which then sink and collide with the VMS. The high densities presented in these clusters, quickly form a VMS in the center that bound several low-mass stars. The new collision treatment implemented in \textsc{MOCCA}, implies that if a star presents a pericenter distance slightly greater than the sum of the radii of its two stars, it is not considered as a candidate to collide with the VMS, however, due to the standard collision treatment, they will be candidates to collide with other stars that present similar conditions (i.e. $r_{\mathrm{p},2} > R_1 + R_2$), thus implying that they might collide with each other, forming a more massive star, that later sink to the center. This extremely dense regime for the star cluster challenges the standard collision treatment in \textsc{MOCCA}, this new update is the first attempt to treat collisions in the proximity of a single massive object at the center, however, the general agreement on the final outcome of both codes is a very good sign that we are on the right track, in the future we will further improve the treatment in \textsc{MOCCA} to obtain better overall agreement for the number of collisions we see with direct \textit{N}-body.


\section{Number of escapers and type of direct collisions in \textit{N}-body} \label{Escapers_bin_and_hyp_Coll}

We analyze the number of escapers, binary and hyperbolic collisions across our models. This analysis is limited to \textit{N}-body simulations.  In \textsc{MOCCA}, all recorded collisions are treated as hyperbolic encounters, since the code does not explicitly follow the orbital motion of stars, thus is not possible to distinguish between binary and hyperbolic collisions.

\begin{figure}[!h]
    \centering
    \includegraphics[]{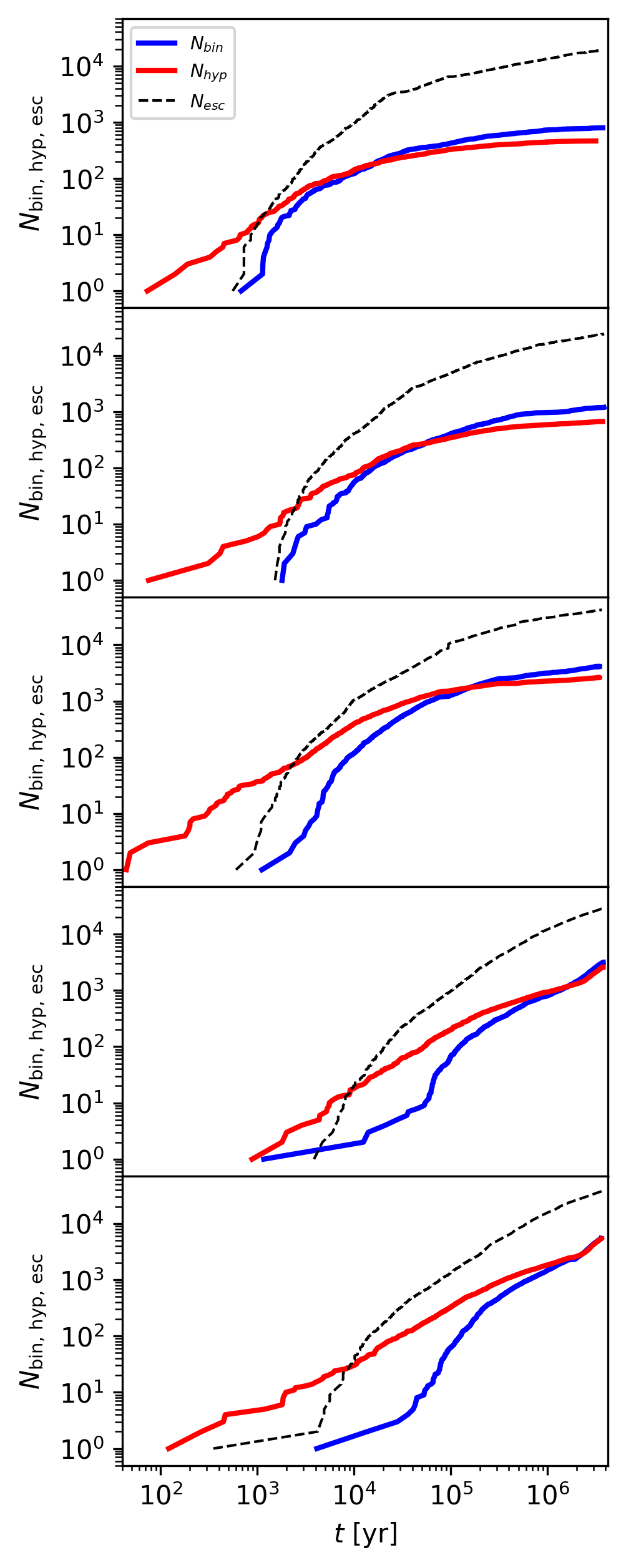} 
    \caption{Time evolution of the number of escaping stars (dashed black line) and the number of binary (solid blue line) and hyperbolic (solid red line) collisions. From top to bottom, the models are \texttt{R0005N50k}, \texttt{R001N100k}, \texttt{R001N250k}, \texttt{R005N500k}, and \texttt{R005N750k}.}
    \label{fig:esc_bin_hyp}
\end{figure}

In Fig.~\ref{fig:esc_bin_hyp}, models \texttt{R0005N50k}, \texttt{R001N100k}, and \texttt{R001N250k} present hyperbolic collisions within the first $100\,$yr, while binary collisions appear later around $1000\,$yr. Notably, the appearance of escaping stars coincides with the onset of binary collisions. Model \texttt{R005N500k}, both hyperbolic and binary collisions occur almost simultaneously at early times. However, hyperbolic encounters dominate for the first few thousand years. Still, the number of escapers begins to rise significantly once the rate of binary collisions increases. Model \texttt{R005N750k}, also shows early hyperbolic collisions, with a single escape occurring prior to any binary collisions. Nevertheless, a notable increase in escapers follows the onset of binary encounters.

In general, we observe that hyperbolic collisions generally occur earlier than binary collisions. The number of escaping stars begins to increase notably as the number of binary collisions rises. This suggests that the stellar escapes are primarily driven by binary interactions. 

\end{appendix}

\end{document}